\documentclass[12pt]{article}
\usepackage{framed}
\usepackage{amsmath}
\numberwithin{equation}{section}
\usepackage{amssymb,amscd}
\usepackage{bm}
\usepackage{wrapfig}
\usepackage{color}
\usepackage{bm}
\usepackage[all]{xy}

\usepackage{theorem}

\newtheorem{theorem}{Theorem}[section]

 \newtheorem{lemma}[theorem]{Lemma}

{\theorembodyfont{\normalfont}

 }
\usepackage{amsbsy} 
\usepackage{theorem}

\def\be{\begin{equation}}
\def\ee{\end{equation}}
\def\bea{\begin{eqnarray}}
\def\eea{\end{eqnarray}}

\newcommand{\rd}{\mathrm{d}}

\def\g{\mathfrak{g}}

\newcommand*{\R}{{\mathbb R}}

\newcommand{\mom}{\mathrm{antifdeg}}

\newcommand{\bx}{\boldsymbol{X}}

\def\ba{\mbox{\boldmath $A$}}

\def\bR{{\mathbb{R}}}

\newcommand{\calH}{{\cal H}}

\newcommand{\calJ}{{\cal J}}
\newcommand{\calL}{{\cal L}}

\newcommand{\calN}{{\cal N}}

\def\xzero{X{}}
\def\tilx{{\mathbb X}{}}
\def\tila{{\mathbb A}{}}

\def\beqa{\begin{eqnarray}}
\def\eeqa{\end{eqnarray}}
\def\beq{\begin{equation}}
\def\eeq{\end{equation}}
\def\be{\begin{equation}}
\def\ee{\end{equation}}
\newcommand{\uX}{{{X}}{}}
\newcommand{\uAplus}{{{A}^+}{}}
\newcommand{\ucplus}{{{c}^+}{}}
\newcommand{\uc}{{{c}}{}}
\newcommand{\uA}{{{A}}{}}
\newcommand{\uXplus}{{{X}^+}{}}

\def\sd{\mbox{\boldmath $\mathrm{d}$}}
\def\bbd{\mbox{\boldmath $\rd$}}

\newcommand\Sbvbbd{{S_{BV}^{(FHGD)}}{}}

\def\pprime{p^{\prime}{}}
\def\xprime{x^{\prime}{}}

\newcommand{\uv}{{\iota_{v}}{}}

\newcommand{\V}{\mathbb{V}}
\newcommand{\W}{\mathbb{W}}
\newcommand{\gdeg}{\g \mathrm{deg}}

\begin{document}


\baselineskip 6mm 

\begin{titlepage}
\begin{flushright}
\end{flushright}


\begin{center}
{\Large \bf From BFV to BV and spacetime covariance}
\vskip 1cm
Noriaki Ikeda${}^{a}$
\footnote{E-mail:\ 
nikedaATse.ritsumei.ac.jp}
and 
Thomas Strobl${}^{b}$
\footnote{E-mail:\ 
stroblATmath.univ-lyon1.fr
}

\vskip 0.4cm
{
\it
${}^a$
Department of Mathematical Sciences,
Ritsumeikan University \\
Kusatsu, Shiga 525-8577, Japan \\

\vskip 0.4cm
${}^b$
Institut Camille Jordan,
Universit\'e Claude Bernard Lyon 1 \\
43 boulevard du 11 novembre 1918, 69622 Villeurbanne cedex,
France

}
\vskip 0.4cm

{
July 24, 2020
}
\vskip 1.6cm
\vskip 1.6cm

\begin{abstract} The BFV formulation of a given gauge theory is usually significantly easier to obtain than its BV formulation. Based on foundational work by Fisch and Henneaux, Grigoriev and Damgaard introduced simple formulas for obtaining the latter from the former. Since BFV relies on the Hamiltonian version of the gauge theory, however, it does not come as a surprise that in general the resulting BV theory does not exhibit space-time covariance. 
%
We provide an explicit example of this phenomenon in two spacetime dimensions and show how to restore covariance of the BV data by improving the Fisch--Henneaux--Grigoriev--Damgaard procedure with 
appropriate adaptations of their formulas.
\end{abstract}
keywords: Gauge symmetry; sigma models; topological field theories;  BV formalism; BFV formalism; AKSZ theories; superspace; supergeometry; Lie algebroids.
\end{center}
\end{titlepage}

\newpage

\tableofcontents
\newpage 
\section{Introduction}
Conventionally, quantum theory on the level of atoms and solid state physics uses Hamiltonian methods. This is in sharp contrast to quantum field theories  (QFTs), which are almost exclusively described by a formalism much closer to Lagrangian methods. One of the main reasons for the latter fact is explicit spacetime covariance: Quantum field theories are defined on $d=(n+1)$-dimensional spacetime manifolds $(\Sigma,\gamma)$ where the pseudo-Riemannian metric $\gamma$ has a Lorentzian signature. In fact, in conventional QFTs as they proved so successful to describe the physics of elementary particles,  $\Sigma = \R^d$, $\gamma$ is a flat Lorentzian metric ($\gamma = \eta$ in the conventional notation), and $d=4$. The physics described by such theories is invariant (or ``covariant'') with respect to the Poincare group $G$, the automorphism group of Minkowski space $(\R^d,\eta)$, and so is the underlying Lagrangian action functional. $G$-invariance then becomes an important restriction to possible counter terms in the renormalization process, for example, and in the dimensional regularization scheme one even admits non-integer values of $d$ so as to keep $G$-covariance inherent at all the intermediary steps. 

A consistent gauge fixing in the context of gauge field theories requires BRST methods \cite{Faddeev-Popov,BRST1,BRST2, Tyutin} or, if one deals with a system that has an ``open algebra'' of gauge symmetry generators, the BV formalism \cite{BV1,BV2}. The construction of the BV extension of a given action functional is in part not always so straightforward, even if existence theorems have been established \cite{Henneaux-Fisch,Henneaux-Teitelboim}. The Hamiltonian framework, on the other hand, offers not only systematic methods to find all the gauge symmetries of a given action functional---something absent in the Lagrangian setting---but also the construction of the BFV formulation \cite{Batalin:1977pb,Batalin:1983pz} of a gauge theory, the Hamiltonian counterpart of its BV formulation, is by experience significantly simpler. It is therefore very useful to have a formalism which yields the BV form of a theory given its BFV data. 

Such a formalism was proposed by Fisch and Henneaux  \cite{Henneaux-Fisch2} (see also \cite{Henneaux-Fisch3}) long ago and improved henceforth by Grigoriev and Damgaard \cite{GrigorievD,Grigoriev}.\footnote{The findings of Grigoriev and Damgaard were inspired  by the superfield formulas introduced previously in a related, but different context by Batalin, Bering and Damgaard \cite{BBD1,BBD2}.} In the Grigoriev--Damgaard reformulation, this formalism can be viewed essentially as the AKSZ method applied to the BFV data of the theory (at the price of an infinite-dimensional target space and, crucial for obtaining a theory with non-vanishing Hamiltonian, by permitting the target data to depend on the odd time of the source; see also section 10 of \cite{Ikeda-Strobl}). One of the strengths of this Fisch--Henneaux--Grigoriev--Damgaard (FHGD) method is that it provides the BV symplectic form and the BV extension of the classical action functional as very concrete and rather simple formulas in terms of the BFV symplectic form, the BFV charge, and the BFV extension of the Hamiltonian. 

What we will address in this article is that this procedure does not always yield a BV action that is (explicitly) $G$-invariant. We will demonstrate this in particular at the example of the twisted Poisson sigma model \cite{Klimcik-Strobl}. This is a topological field theory in two spacetime dimensions. Correspondingly, in this case the group $G$ consists of all diffeomorphisms of $\Sigma$, $G= \mathrm{Diff}(\Sigma)$. The FHGD method produces formulas in which only the subgroup of spatial diffeomorphisms $G_{sp} = \mathrm{Diff}(\Sigma_1)$ is preserved inherently (within the Hamiltonian formalism one  restricts spacetime to the form $\Sigma := \Sigma_1 \times \R$). As argued above, one of the main reasons for the use of Lagrangian and BV methods is precisely the covariance or $G$-invariance within the process of quantization. We will show here, how to restore it in the formulas one obtains for the  twisted Poisson sigma model and how to approach this problem in the general case.

There are essentially two independent potential reasons for the deficiency of $G$-invariance.  The first one arises in the context of Wess-Zumino (WZ) terms: Adding the pullback of a closed ($d+1$)-form $\varphi$  to the action functional, which then becomes non-local, the BFV symplectic form $\omega_{BFV}$ remains local, but it is no more exact. In particular, it therefore cannot be cast into Darboux form globally. The $\varphi$-contribution to  $\omega_{BFV}$ spoils simultaneously the Darboux form \emph{and} the $G$-covariance of the resulting BV symplectic form $\omega_{BV}^{(FHGD)}$. Now, 
for degree reasons, every $BV$-symplectic form is exact, so also this one. However, one needs to be careful in this context, as we will illustrate by means of examples: when restoring the Darboux form by a local transformation, it can easily happen that the ghost degree zero part of the BV action does no more coincide with the classical action (which then needs to be restored subsequently by a further local BV-canonical redefinition of fields).

The second reason is less evident: In the Hamiltonian formulation of a gauge theory, it is the constraints which generate the gauge transformations of the fields on phase space. These transformations are uniquely defined on all of phase space, but in principle one can add to them contributions that vanish on the constraint surface, while not spoiling the property to be a  symmetry of the theory. For gauge invariance of the corresponding Hamiltonian action functional $S_{Ham}$ with respect to a given transformation of the canonical fields, the Lagrange multiplier fields, which enforce the constraints within $S_{Ham}$, need to transform appropriately. Spacetime covariance of the generators of (non-trivial) gauge transformations may now require a balanced interplay between the transformation properties of the phase space variables and the Lagrange multipliers. In general this requires appropriate additions to the transformation formulas for the phase space variables of terms vanishing on the constraint surface. In the FHGD formalism in its current form such terms are always absent.

This is of relevance here since the gauge transformation of the fields can be read off from the part of the BV action which is linear in the antifields. If these transformations are not $G$-covariant, the corresponding BV action is not as well. We will show in this paper that the covariance of gauge transformations can be always restored in the BV action by a BV canonical transformation. At least for the Poisson sigma model twisted by a WZ term, the implementation of the combination of the two field redefinitions---the one that brings $\omega_{BV}^{(FHGD)}$ into Darboux form (without changing the classical part of the BV action) followed by a BV symplectic transformation which leads to a covariant part of the BV action linear in the antifields---suffices to ensure covariance of all of the BV extension obtained in this way.

The structure of the paper is as follows: In Section \ref{secFHGD}, we recall the main formulas from the FHGD formalism. Since in our application to the twisted Poisson sigma model, both the BFV and the BV formulation of the theory admit a more concise description in terms of superfields, we adapt the FHGD formalism to include such a formulation. Likewise for the inclusion of WZ terms. 
 In Section \ref{secHPSM}, we recall the main facts about the two-dimensional sigma model based on our recent article \cite{Ikeda-Strobl2} and apply the FHGD formalism to it. 
As announced, this will lead to non-covariant formulas for both $\omega_{BV}^{(FHGD)}$ and $S_{BV}^{(FHGD)}$. 

In Section \ref{secToy} we consider mechanical toy models with Lie algebroid gauge symmetries.  These models may be of interest also in their own right. We use them to clarify the source of the deficiency as outlined above and show how to overcome them in each case.  As an important tool in this context, we will provide a lemma which shows that \emph{every} change of the generators of gauge transformations can be implemented in terms of a BV symplectic transformation. In Section \ref{secRestore} we then apply these findings to the twisted Poisson sigma model, recovering the completely covariant BV form of this model constructed recently ab initio in \cite{Ikeda-Strobl2}. In the final section, Section \ref{secGeneral}, we generalize and formalize the improvement procedure: We introduce a double complex governing $G$-invariant extensions, for some group $G$ acting on field space, and provide a recursive cohomological procedure so as  to improve an initially given, not $G$-invariant BV extension to one that has the desired property.


A final word on the use of the term ``covariant'', which, as also pointed out by the anonymous referee, is often used in a not very precise sense in the literature. In a first approximation (and close in spirit to the introductory remarks above about renormalization etc), in the present context one may think of (manifest) spacetime covariance as that in all formulas upper and lower spacetime indices are always contracted. Section \ref{secGeneral} provides an alternative, more mathematically suitable definition, where this property is expressed as the closedness with respect to an appropriate differential.

\section[Generalities of the Fisch--Henneaux--Grigoriev--Damgaard formalism]{Generalities of the FHGD formalism}
\label{secFHGD}
In this section, we briefly review the  procedure to construct a BV symplectic form and the BV extension of a gauge theory when starting its Hamiltonian BFV formulation. This formalism was initiated by Fisch--Henneaux and improved by Grigoriev--Damgaard, for which reason we call it the  Fisch--Henneaux--Grigoriev--Damgaard (FHGD) formalism.  
We will not refrain from using rather explicit formulas in this section (as well as in the following ones) and extend the already improved setting of Grigoriev and Damgaard so as to include BFV superfields and WZ terms.

\subsection{Traditional setting}\label{traditional}

Let $\Sigma = \Sigma_n \times \bR$ denote the $d=(n+1)$-dimensional spacetime manifold of a given field theory, equipped with local coordinates $\sigma^{\mu}$, $\mu= 0, 1, \ldots, n$, where $x ^0 \in \bR$ denotes the evolution parameter or time coordinate. 

The FHGD formalism sets in after having constructed the BFV form of a constrained system, here for some fields on $\Sigma_{n}$. Let $z^I(\sigma)$ denote these fields, including the ghosts. The BFV data are then given by a BFV symplectic form $\omega_{BFV}$, an odd function $S_{BFV}(\sigma)$, which is BFV-BRST charge, and an even function $H_{BFV}(\sigma)$, which is a BFV extension of the Hamiltonian functional. Since we deal with local field theories, the BFV symplectic form can be written as 
\begin{eqnarray}
\omega_{BFV} = \int_{\Sigma_n} d^n \sigma \,
\omega_{IJ}(z) \delta z^I \wedge \delta z^J,
\label{BFVsymplectic}
\end{eqnarray}
where $\omega_{IJ}(z)$ is a nondegenerate, graded-antisymmetric matrix and $\delta$ denotes the de Rham differential on field space.  Likewise, the charge $S_{BFV}$ and the Hamiltonian $H_{BFV}$ are
integrations of densities,
\begin{eqnarray}
S_{BFV} &=& \int_{\Sigma_n} d^n \sigma \, {\cal J}(z(\sigma)),
\\
H_{BFV} &=& \int_{\Sigma_n} d^n \sigma \, {\cal H}(z(\sigma)),
\end{eqnarray}
for some functions ${\cal J}$ and ${\cal H}$. In the above, ${\cal J}$ and ${\cal H}$ can depend on 
$z(\sigma)$ and its derivatives $\partial z, \ldots$ (up to some finite order); they are functions on the jet bundle. For notational simplicity we will simply write  ${\cal J}(z)$ and ${\cal H}(z)$ for these functions in what follows, but with this interpretation.

$S_{BFV}$ and $H_{BFV}$ satisfy the following equations with  respect to the graded Poisson brackets induced by $\omega_{BFV}$:
\begin{eqnarray}
&& \{S_{BFV}, S_{BFV} \} =0,
\label{BFVequation1}
\\
&& \{S_{BFV}, H_{BFV} \} =0,
\label{BFVequation2}
\\
&& \{H_{BFV}, H_{BFV} \} =0.
\label{BFVequation3}
\end{eqnarray}

The FHGD procedure to construct the BV data, in the formulation of Grigoriev and Damgaard,  then works as follows. First, for each field $z^I(\sigma)$ one introduces a superpartner field $w^I(\sigma)$. In addition, it is convenient to introduce a superpartner coordinate $\theta^0$ corresponding to time  $t \equiv \sigma^0$. This permits one to introduce a superfield $Z^I(\sigma, \theta^0)$ by means of
\begin{eqnarray}
Z^I(\sigma, \theta^0) = z^I(\sigma) + \theta^0 w^I(\sigma). \label{Z}
\end{eqnarray}
The BV symplectic form just becomes a super extension of 
the BFV symplectic form \eqref{BFVsymplectic},
\begin{eqnarray}
\omega_{BV} = \int_{\Sigma_n \times T[1]\bR} \!\!\!\! d \theta^0 dt d^n \sigma \, \omega_{IJ}(Z) \delta Z^I \wedge \delta Z^J,
\label{BVsymplectic}
\end{eqnarray}
where $\delta$ now denotes the de Rham differential on the extended BV space of fields.

In the traditional FHGD procedure one needs to add a further assumption to the BFV data so as to construct the BV action functional: The BFV symplectic form should be exact, $\omega_{BFV} = - \delta \vartheta_{BFV}$ for some local $1$-form \begin{eqnarray} \label{thetaBFV}
\vartheta_{BFV} &=& \int_{\Sigma_n} \!\!\!\! d^n \sigma \, \vartheta_I(z) \delta z^I,
\end{eqnarray}
parametrized by $\vartheta_I(z)$ and one is given this 1-form also. 
%
%
%
Then one can define the BV action $\Sbvbbd$ as follows:
\begin{align}
\Sbvbbd
&:= \int_{\Sigma_n \times T[1]\bR} 
\!\!\!\! d \theta^0 d t d^n \sigma \, \vartheta_I(Z) \bbd_0 Z^I
- \int_{T[1]\bR} \!\!\!\! d \theta^0 d t \, (S_{BFV}(Z)+ \theta^0 H_{BFV}(Z))
\nonumber \\
&= \int_{\Sigma_n \times T[1]\bR} 
\!\!\!\! d \theta^0 d t d^n \sigma \, 
(\vartheta_I(Z) \bbd_0 Z^I - \calJ(Z) - \theta^0 \calH(Z)).
\label{FHGDBVaction}
\end{align}
Here $\bbd_0 \equiv \theta^0 \partial_0$ can be viewed as the de Rham differential on the line $\R$ or, better, the corresponding odd and nilpotent vector field on its superextension $T[1]\R$. If one assumes that the original coordinates have Darboux form, $z = (q^i,p_i)$, then, correspondingly, $Z = (Q^i, P_j)$ and the first term above can be rewritten as follows:
\begin{eqnarray}
\int_{\Sigma_n \times T[1]\bR} \!\!\!\! d \theta^0 d t d^n \sigma \, \vartheta_I(Z) \bbd_0 Z^I
&=& 
\int_{\Sigma_n \times T[1]\bR} \!\!\!\! d \theta^0 d t d^n \sigma \, P_i \bbd_0 Q^i.
\end{eqnarray}

After integrating out the odd variable $\theta^0$, \eqref{FHGDBVaction} becomes a functional 
for the fields on the spacetime $\Sigma$. 
It satisfies the classical master equation, $(\Sbvbbd,\Sbvbbd) = 0$, as follows 
from equations \eqref{BFVequation1}--\eqref{BFVequation3}.
Here $(-,-)$ is the BV bracket induced from the BV symplectic form
\eqref{BVsymplectic}. Also, by construction, the ghost degree zero part of $\Sbvbbd$ agrees with the Hamiltonian action functional of the original gauge theory, which, by the equivalence of Lagrangian and Hamiltonian methods, can be assumed to be equivalent to the original action functional (possibly after integrating out some momenta, for example). Likewise, also all gauge symmetry generators are included in the ghost sector of \eqref{FHGDBVaction} so that one can view $\Sbvbbd$ together with \eqref{BVsymplectic} and \eqref{Z} as a valid BV description (if one prefers, after integrating out the odd time variable $\theta^0$).

\subsection{Including superfields of the underlying BFV theory}
\label{secsuper}

There exist some topological gauge theories for which the BFV theory can be formulated in terms of superfields, such as AKSZ-sigma models \cite{AKSZ} (see also \cite{CFAKSZ,RoytenbergAKSZ,Ikeda:2012pv}) or the twisted Poisson sigma model \cite{Klimcik-Strobl} (see, in particular, \cite{Ikeda-Strobl2}). In this case the BFV fields combine into superfields 
${\mathbb Z}^I(\sigma,\theta)$ defined on $\calN = T[1]\Sigma_n$ and depending on coordinates $(\sigma^\alpha,\theta^\alpha)_{\alpha=1}^n$. Certainly, such theories are in principle already covered by the formulas above---one just needs to integrate out the odd spatial variables $\theta^\alpha$ for this purpose. However, it is useful in such cases, to not do so, but to combine the spatial BFV superfields 
into BV superfields $\mathbf{Z}^I(\sigma,\theta)$, defined on $T[1]\Sigma_n \times T[1]\R \cong T[1](\Sigma_n \times \R)$ and now depending on coordinates $(\sigma^\mu,\theta^\mu)_{\mu=0}^{n}$.  This allows for a much more compact notation in such cases. 

Although the formulas are essentially identical to those in the previous subsection, with rather obvious replacements, for the convenience of the reader and clarity of the presentation we provide the main formulas also in this notation. 

We assume that the BFV data are given in the following form: 
\begin{eqnarray}
S_{BFV} &=& \int_{T[1]\Sigma_n} d^n \sigma d^n \theta \, {\cal J}({\mathbb Z}(\sigma, \theta)),
\\
H_{BFV} &=& \int_{T[1]\Sigma_n} d^n \sigma  d^n \theta \, {\cal H}({\mathbb Z}(\sigma, \theta)).
\end{eqnarray}
In addition,\footnote{Also on the space of fields, we now consider differential forms as functions on its shifted cotangent bundle. Therefore, the wedge product is replaced by a product of functions here.}  
\begin{eqnarray}
\omega_{BFV} = \int_{T[1]\Sigma_n} d^n \sigma d^n \theta\,
\omega_{IJ}({\mathbb Z}) \delta {\mathbb Z}^I \delta {\mathbb Z}^J.
\label{BFVsymplecticsuper}
\end{eqnarray}
where the product of $\delta {\mathbb Z}^I$ and $\delta {\mathbb Z}^J$ becomes a graded product ($\delta$ has degree one in the superfield formalism, for example, but also the fields ${\mathbb Z}^I$ can have some fixed non-zero degree for a fixed choice of the index $I$). As before, we assume the existence of 
\begin{eqnarray}\label{thetaBFVsuper}
\vartheta_{BFV} &=& \int_{T[1]\Sigma_n} \!\!\!\! d^n \sigma d^n \theta \, \vartheta_I({\mathbb Z}) \delta {\mathbb Z}^I,
\end{eqnarray}
such that $\omega_{BFV} = - \delta \vartheta_{BFV}$.

In the FHGD formalism one now introduces superpartners ${\mathbb W}^I(\sigma, \theta)$ for ${\mathbb Z}^I(\sigma, \theta)$, which allows for the definition of the total or BV superfield 
\begin{eqnarray} \label{total}
\mathbf{Z}^I(\sigma, \theta, t, \theta^0) 
= {\mathbb Z}^I + \theta^0 {\mathbb W}^I.
\end{eqnarray}
The BV symplectic form then becomes
\begin{eqnarray}
\omega_{BV} = \int_{T[1](\Sigma_n \times \bR)} \!\!\!\! d \theta^0 dt d^n \sigma d^n \theta \, \omega_{IJ}(\mathbf{Z}) \delta \mathbf{Z}^I \delta \mathbf{Z}^J. \label{omegaBVtotal}
\end{eqnarray}
It is written as an integral over the $(n+1)$-dimensional super manifold given by odd tangent bundle over spacetime $\Sigma = \Sigma_n \times \bR$. Likewise,  the BV functional takes the form:
\begin{align}
\Sbvbbd
&= \int_{T[1]\Sigma_n \times T[1]\bR} 
\!\!\!\! d \theta^0 d t d^n \sigma d^n \theta \, \vartheta_I(\mathbf{Z}) \bbd_0 \mathbf{Z}^I
- \int_{T[1]\bR} \!\!\!\! d \theta^0 d t \, (S_{BFV}(\mathbf{Z})+ \theta^0 H_{BFV}(\mathbf{Z}))
\nonumber \\
&= \int_{T[1]\Sigma}
\!\!\!\! d^{n+1} \sigma d^{n+1} \theta \, 
(\vartheta_I(\mathbf{Z}) \bbd_0 \mathbf{Z}^I - \calJ(\mathbf{Z}) - \theta^0 \calH(\mathbf{Z})).
\label{BVactionsuper}
\end{align}


\subsection{Extension to Wess-Zumino terms}\label{secWZ}
If the action has a WZ term, the BFV symplectic form $\omega_{BFV}$ is necessarily non-exact. Correspondingly, the Liouville 1-form $\vartheta_{BFV}$, which enters the formula \eqref{FHGDBVaction} does not exist, at least not globally, and the FHGD 
procedure needs to be adapted.

When one has a WZ term in the classical action, spacetime $\Sigma$ needs to 
be such that it can be the boundary of another manifold $N$, $\partial N = \Sigma$.
Correspondingly, $\Sigma$ cannot have any boundary itself, $\partial \Sigma =0$.
On the other hand, in the Hamiltonian formulation uses a decomposition of spacetime into 
``space'' and ``time''. To not run into a contradiction with the fact that $\Sigma$ is boundariless, we compactify time to a circle such that 
\beq 
\Sigma = \Sigma_n \times S^1,
\eeq 
and require $\partial 
\Sigma_n =0$. For the manifold $N$ one now has essentially two options, to replace $S^1$ by a disc $D$ or, if $n \neq 0$, to choose some  $N_{n+1}$ such that  $\Sigma_n= \partial N_{n+1}$. In the second case, we still have a time-like direction singled out, which will make the formulas below easier. We will thus choose the second option.

With $\omega_{BFV}$ still being given by \eqref{BFVsymplectic}, there is no problem in adapting 
\eqref{BVsymplectic} to this setting,
\begin{eqnarray}
\omega_{BV} = \int_{\Sigma_n \times T[1]S^1} \!\!\!\! d \theta^0 dt d^n \sigma \, \omega_{IJ}(Z) \delta Z^I \wedge \delta Z^J.
\label{BVsymplecticwithS1}
\end{eqnarray}
Instead of \eqref{FHGDBVaction}, on the other hand, we propose the following adaptation to WZ terms:
\begin{align}
\Sbvbbd
&= \int_{N_{n+1} \times T[1]S^1} \!\!\!\!\!\!\!\! 
d \theta^0 dt d^n \sigma \, \rd_{n+1} Z^I \omega_{IJ}(Z) \bbd_0 Z^J 
- \int_{\Sigma_n \times T[1]S^1} 
\!\!\!\! d \theta^0 d t d^n \sigma \, 
(\calJ(Z) + \theta^0 \calH(Z)), \label{FHGDWZ}
\end{align}
where $\rd_{n+1}$ is the de Rham differential on $\Sigma_{n+1}$. Analogously, in the case of BFV superfields, we replace \eqref{BVactionsuper} by
\begin{align}
\Sbvbbd
&= \int_{T[1](N_{n+1} \times S^1)} \!\!\!\!\!\!\!\! d^{n+2} \sigma d^{n+2} \theta \, \bbd_{n+1} {\mathbf Z}^I \omega_{IJ}({\mathbf Z}) \bbd_0 {\mathbf Z}^J
- \int_{T[1]\Sigma} \!\!\!\! d^{n+1} \sigma d^{n+1} \theta \, 
(\calJ({\mathbf Z}) + \theta^0 \calH({\mathbf Z})),
\label{FHGDactionWZsuper}
\end{align}
where $\bbd_0$ and $\bbd_{n+1}$ are superderivatives on $T[1]S^1$ and $T[1]N_{n+1}$, respectively.

We conclude this section with a final remark on the topology of $\Sigma$ within the FHGD procedure in general: To retrieve the BV formulation of the original action from $\Sbvbbd$ in \eqref{BVactionsuper}, one will forget at the end that $\Sigma = \Sigma_n \times \R$ (or likewise $\Sigma = \Sigma_n \times S^1$) was used in the intermediary steps and generalize the result to arbitrary $d=(n+1)$-dimensional spacetime manifolds $\Sigma$ (and possibly integrate out some additional fields used in the first order formulation that might not have been present in the original action functional). A similar statement holds true for what concerns \eqref{FHGDWZ} and \eqref{FHGDactionWZsuper}.


\section{Applying the FHGD method to the twisted Poisson sigma model}
\label{FHGDTPSM}
In this section we present our main example for the covariance problems of the Fisch--Henneaux--Grigoriev--Damgaard procedure when applied to a field theory, the twisted Poisson sigma model \cite{Klimcik-Strobl}. We will first briefly recollect the most elementary basic facts about this two-dimensional topological field theory as well as the BFV data as they were obtained in a previous paper \cite{Ikeda-Strobl2}. We then apply the FHGD method to these data. We will see that the resulting formulas treat ``space'' and ``time'' in a very different fashion and are in particular not explicitly covariant with respect to the initially present diffeomorphism invariance on $\Sigma$. We will contrast this with the explicitly $G=\mathrm{Diff}(\Sigma)$-invariant BV data as they were found in \cite{Ikeda-Strobl2} (but with more work than by applying the FHGD formalism here). In the remaining part of this article we will then show how the non-covariant BV data of the FHGD formalism can be transformed into these $G$-covariant ones.

\label{secHPSM}
\subsection{Classical action and its gauge symmetries}
A twisted Poisson manifold \cite{Park,Klimcik-Strobl,Severa-Weinstein} is a manifold $M$ equipped with a bivector field $\pi \in \Gamma(\wedge^2 TM)$  and a closed 3-form $H \in \Omega^3(M)$ such that 
\begin{equation} \label{PiPi}
 \tfrac{1}{2}[\pi,\pi] = \pi^\sharp(H) \, ,
\end{equation}
where the bracket on the left-hand side denotes the Schouten bracket of multivector fields and $\pi^\sharp$ denotes the canonical map from $T^*M$ to $TM$, applied to each factor of $\wedge^3 T^*M$ on the right-hand side.\footnote{Since the twisted Poisson sigma model is topological and has an on-shell vanishing Hamiltonian, we can use the conventional letter $H$ for the closed 3-form in this context without running into danger of confusion.}

The $H$-twisted Poisson sigma model (HPSM) is defined on a three-dimensional manifold $N$ with boundary $\Sigma = \partial N$, the target space being a twisted Poisson manifold $(M, \pi, H)$.
For concreteness we will choose $\Sigma$ and $N$ to be a torus and  a solid torus, respectively,
\beq \Sigma = S^1 \times S^1 \quad , \qquad N = D \times S^1 \, ,  \label{torus}
\eeq 
where $D$ denotes a two-dimensional disc. The classical action $S$ is defined on the space of maps $X \colon N \rightarrow M$ together with the fields $A \in \Omega^1(\Sigma, X^*T^*M)$ and takes the form
\begin{eqnarray} 
\! \!\!S = \int_{\Sigma=\partial N} 
A_i \wedge \rd \xzero^i + \tfrac{1}{2} \pi^{ij}(X) A_i \wedge A_j 
+ \int_{N} X^*H \, ,
\label{classicalactionofHPSM}
\end{eqnarray}
where the last term is a WZ-term. In the case that $H$ is exact, $H=\rd B$, we obtain a $B$-twisted 
Poisson sigma model (BPSM), which is an inherently local, two-dimensional field theory, defined for every orientable $\Sigma$:
\begin{eqnarray}
\! \!\!S = \int_{\Sigma} 
A_i \wedge \rd \xzero^i + \tfrac{1}{2} \pi^{ij}(X) A_i \wedge A_j 
+  \tfrac{1}{2} B_{ij}(X) \rd \xzero^i \wedge \rd \xzero^j \, .
\label{BPSM}
\end{eqnarray}
In the particular case, where even $H=B=0$, this theory reduces to the ordinary Poisson sigma model (PSM) \cite{Ikeda-Izawa,Ikeda,Schaller-Strobl,Schaller-Strobl2}, which coincides with the AKSZ-theory \cite{AKSZ} in two dimensions then.

The Euler-Lagrange equations of the HPSM (or also the BPSM) take the following form,
\begin{eqnarray}
F^i := \rd X^i + \pi^{ij} A_j &=& 0  \, ,
 \label{eomdX}\\
 \rd A_i + \tfrac{1}{2} \pi^{jk},_{i}
A_j \wedge A_k + \tfrac{1}{2} H_{ijk} \rd X^j\wedge \rd X^k &=&0  \, .
\label{eomdA}
\end{eqnarray}
For the 1-form field equations we introduced the letter $F$ as an abbreviation, since they will play an important role later on. In particular, it turns out (see below) that the first class constraints of this model coincide with the 1-form part of $F^i$, i.e. $F_1^i = \partial_1 X^i + \pi^{ij} A_{1j} \approx 0$.

Generators of the gauge transformations can be chosen to be of the form 
\begin{eqnarray}
\delta_\epsilon \xzero^i &=& - \pi^{ij} \epsilon_j,
\label{trafoX}
\\
\delta_\epsilon A_{i}  
&=& 
\rd \epsilon_i + f_i^{jk} A_{j} \epsilon_k
+ \tfrac{1}{2} \pi^{jk} H_{ijl} \, F^j \, \epsilon_k,
\label{gaugetransformation02a}
\end{eqnarray}
where $\epsilon_i$ is a gauge parameter and
\beq f_i^{jk} \equiv \pi^{jk}{},_i + \pi^{jl} \pi^{km} H_{ilm} . \label{fijk} \eeq
In fact, the above generators can be put into such a form only if the field $X$ maps into a single patch  $U \subset M$ of coordinates. One option for a \emph{target space covariant} presentation of the gauge symmetries is the following one:
\begin{eqnarray}
\delta^{\nabla}_\epsilon \xzero^i &=& - \pi^{ij} \epsilon_j,
\label{trafoXnabla}
\\
\delta^\nabla_\epsilon \! A_{i} &=& \rd \epsilon_i
+ f_i^{jk} A_{j} \epsilon_k
- \Gamma^k_{ij} \, F^j \, \epsilon_k. \label{deltaAnabla}
\end{eqnarray}
Here $\nabla$ is an auxiliary connection $\nabla$ on $M$ with torsion \beq 
\Theta = \langle \pi, H \rangle,  \label{torsionpiH}
\eeq 
whose connection coefficients are denoted by $\Gamma^k_{ij}$. \eqref{gaugetransformation02a} follows from the above  \eqref{deltaAnabla} for the choice of connection and coordinates on $M$ such that the symmetric part of the coefficients  $\Gamma^k_{ij}$ vanish. 

We discussed the target space covariance of the theory at length in \cite{Ikeda-Strobl2} and in the present paper will content ourselves mostly with the local representatives \eqref{trafoX} and \eqref{gaugetransformation02a}.\footnote{It turns out, however, that the superfield expressions, also the ones obtained by the FHGD method below, will have some global significance on the target space, see \cite{Ikeda-Strobl2} for the corresponding transformation behavior of the fields.} What is important for the present context is that these generators are evidently covariant with respect to 
\beq G = \mathrm{Diff}(\Sigma) . \label{Diff} \eeq
One verifies by an explicit, albeit somewhat lengthy calculation that they form an open algebra (this is true for generic choices of $\pi$ and $H$). Thus, for the gauge fixing and a subsequent quantization, the BV formalism is mandatory.

\subsection{The Hamiltonian and the BFV formulation}\label{BFVHPSM}
In this subsection we mainly summarize results about the BFV formalism of the HPSM as they can be found in \cite{Ikeda-Strobl2}.

For the Hamiltonian treatment, we declare one of the two factors $S^1$ of $\Sigma$ in \eqref{torus} to correspond to ``space'' (parametrized by the coordinate $\sigma^1$) and the other one to ``time''
(coordinate $\sigma^0$). The symplectic form is
\begin{eqnarray}
\omega &=& \oint_{S^1}  \rd \sigma^1 
\left(\delta \xzero^i \wedge \delta A_{1i}
+ \tfrac{1}{2} H_{ijk} \, \partial_1 \xzero^i  \,
\delta \xzero^j \wedge \delta \xzero^k
\right)\, ,
\label{classicalsymplecticform}
\end{eqnarray}
where $\partial_1 \equiv \partial/\partial \sigma^1$ and all fields depend on $\sigma^1$, giving  rise to the following Poisson brackets: 
\begin{eqnarray}
\{\xzero^i(\sigma), \xzero^{j}(\sigma^{\prime})\}  &=&0, \nonumber \\
\{\xzero^i(\sigma), A_{1j}(\sigma^{\prime})\}  &=& \delta^i_j \delta(\sigma - \sigma^{\prime}),
\label{classicalPB}
\\
\{A_{1i}(\sigma), A_{1j}(\sigma^{\prime})\}  &=& 
- H_{ijk}(\xzero) \partial_1 \xzero^k \, \delta(\sigma - \sigma^{\prime}). \nonumber
\end{eqnarray}
The ``time''-components $A_{0i}$  of the fields $A_i$, on the other hand, serve as Lagrange multipliers for the following constraints 
\begin{eqnarray} \label{G}
J^i \equiv \partial_1 \xzero^i + \pi^{ij}(\xzero) A_{1j} \approx 0.
\end{eqnarray}
They are of the first class since their Poisson brackets close among one another:
\begin{eqnarray} \label{constraintalgebra}
\{J^i(\sigma), J^j(\sigma^{\prime})\}  = - 
f^{ij}_k(X(\sigma)) \, J^k(\sigma) \delta(\sigma - \sigma^{\prime})\, ,
\end{eqnarray}
where the functions $f^{ij}_k(x)$ were introduced in \eqref{fijk}; they are the structural functions of the Lie algebroid on $T^*M$ as induced by the twisted Poisson structure. As for every $\mathrm{Diff}(\Sigma)$-invariant theory, the Hamiltonian vanishes on-shell,
\begin{eqnarray}
\mathrm{Ham} =  \oint_{S^1} \rd \sigma A_{0i} J^i  \approx 0,
\label{TPSMhamiltonian}
\end{eqnarray}
and does not play any important role here.


In order to construct the BFV formalism, one introduces a ghost pair 
$c^i(\sigma)$ and $b_i(\sigma)$ of ghost number $1$ and $-1$, respectively, which Poisson commute with the previous fields and satisfy the following Poisson bracket among one another,
\begin{eqnarray}
\{c_i(\sigma), 
b^j(\sigma^{\prime}) \}&=& 
\delta_i^j \delta(\sigma - \sigma^{\prime}).
\end{eqnarray}
This corresponds to the following BFV symplectic form:
\begin{eqnarray}
\omega_{BFV} = \oint_{S^1} \rd \sigma^1
\left(\delta \xzero^i \wedge \delta A_{1i}
+ \delta c_i \wedge \delta b^i
+ \tfrac{1}{2} H_{ijk}(\xzero) \partial_1 \xzero^i \,
\delta \xzero^j \wedge \delta \xzero^k \right).
\label{BFVsymplecticform}
\end{eqnarray}
The odd BFV functional takes the minimal form
\beq \label{SBFVfield} 
S_{BFV} = \oint_{S^1} \rd \sigma^1 \left(
c_i\,J^i +\tfrac{1}{2} f^{ij}_k \,  c_i c_j b^k \right),
\eeq
and satisfies $ 
\{ S_{BFV},S_{BFV}\} = 0$.  

These data permit a superfield reformulation: Introducing the odd coordinate $\theta^1$, to which we assign ghost number one, the BFV fields can be recombined into
\begin{eqnarray}
&& \tilx^i (\sigma^1, \theta^1) := \xzero^i (\sigma^1)+ \theta^1 \, b^i (\sigma^1),
\label{superfield1dx} \\
&& \tila_i (\sigma^1, \theta^1) := - c_i (\sigma^1) + \theta^1 \, p_{ i} (\sigma^1),
\label{superfield1da}
\end{eqnarray}
which are now superfields of degree $0$ and $1$, respectively. Then the BFV symplectic form \eqref{BFVsymplecticform} becomes
\begin{eqnarray}
\omega_{BFV} = \int_{T[1]S^1} \rd \sigma^1 \rd \theta^1
\left(\delta \tilx^i  \delta \tila_{i}
+ \tfrac{1}{2} H_{ijk}(\tilx) {\bbd}_1 \tilx^i \,
\delta \tilx^j  \delta \tilx^k \right),
\label{superBFVsymplecticform}
\end{eqnarray}
where ${\bbd}_1 = \theta^1 \tfrac{\partial}{\partial \sigma^1}$ is the odd vector field on $T[1]S^1$ corresponding to the de Rham differential on $S^1$. Denoting by ${\bm{\varepsilon}}_1$ the Euler vector field on this supermanifold, 
${\bm{\varepsilon}}_1 = \theta^1 \frac{\partial}{\partial \theta^1}$, 
 the BFV functional \eqref{SBFVfield} is rewritten as
\begin{eqnarray}
S_{BFV}
&=&
\int_{T[1]S^1} \!\!\!\!\!\!\!\! \rd \sigma^1 \rd \theta^1
\left(\tila_i \, {\bbd}_1 \tilx^i
+ \tfrac{1}{2} \pi^{ij}(\tilx) \tila_i \tila_j
+ \tfrac{1}{2} \pi^{il} \pi^{jm} H_{klm}(\tilx)\,
\tila_i \tila_j \, {\bm{\varepsilon}}_1\tilx^{k} 
\right).
\label{superBFVBRSTcharge}
\end{eqnarray}
The BFV symplectic form \eqref{superBFVsymplecticform} is not exact for a non-exact 3-form $H$. In the case of a $B$-twisted Poisson sigma model, on the other hand, $H=\rd B$, one can bring $\omega_{BFV}$ into Darboux form by introducing the canonical momentum 
\beq p_i = A_{1i} - B_{ij}(X) \partial_1 X^j \, . \label{3.25}
\eeq
If $M=\R^{\dim(M)}$ and $x^i$ cartesian coordinates, then globally
\begin{eqnarray}
\omega_{BFV} = \oint_{S^1} \rd \sigma^1
\left(\delta \xzero^i \wedge \delta p_{i}
+ \delta c_i \wedge \delta b^i \right).
\label{BFVsymplecticformXp}
\end{eqnarray}
$S_{BFV}$ still is of the form \eqref{superBFVBRSTcharge}, but with the constraints 
\beq  J^i = \partial_1 \xzero^i + \pi^{ij}(\xzero) p_{j} + \pi^{ij} B_{jk} \, \partial_1 X^k. \label{JB}
\eeq


\subsection[BV from the Fisch--Henneaux--Grigoriev--Damgaard formalism for the HPSM]{BV from the FHGD formalism for the HPSM}\label{secBVFHGDHPSM}
In this subsection we finally apply the FHGD formalism, reviewed and slightly extended in Section \ref{secFHGD}, to the BFV form of the HPSM recollected in Section \ref{BFVHPSM}. In particular, it turns out useful to use directly the superfield formalism of the BFV-HPSM. Applying the general strategy of Section \ref{secsuper}, we introduce superpartner fields for each of the fundamental fields $\tilx^i$ and $\tila_i$, which we denote by $\tila\!^{+i}$ and $\tilx_{i}^+$, respectively. For later identifications and a comparison with the $G$-covariant BV formalism, it will be useful to parametrize them as follows (as fields on $T[1]S^1$)
\begin{eqnarray}
&& \tila\!^{+i}(\sigma^1, \theta^1) := - A_0^{+i} (\sigma^1) - \theta^1 c_{10}^{+i} (\sigma^1), \label{a0}
\\
&& \tilx_{i}^+ (\sigma^1, \theta^1) 
:= A_{0 i} (\sigma^1) - \theta^1 \xzero_{10 i}^{+} (\sigma^1). \label{x0}
\end{eqnarray}
According to \eqref{total}, these four fields  combine into the BV superfields 
\bea 
\bx^i &:= & \tilx^i + \theta^0 \, \tila\!^{+i}, \label{bx}\\ 
\ba_i &:=&\tila_i + \theta^0  \, \tilx_{i}^+ .\label{ba}
\eea 
They are now considered as functions on $T[1](S^1 \times S^1)= T[1]\Sigma$, so they depend on  $(\sigma, \theta)$, which is short for $( \sigma^0, \sigma^1, \theta^0, \theta^1)$. Altogether, substituting \eqref{superfield1dx}, \eqref{superfield1da}, \eqref{a0}, and \eqref{x0} into \eqref{bx} and \eqref{ba}, we obtain
\begin{eqnarray}
\bx^i (\sigma, \theta)
&= & \uX^i (\sigma) - \uAplus^i (\sigma, \theta)
+ \ucplus^i (\sigma, \theta) ,
\label{2Dsuperfield1} \\
\ba_i (\sigma, \theta) 
&= &
- \uc_i (\sigma) + \uA_i (\sigma, \theta)
+ \uXplus\!\!\!\!_i \,(\sigma, \theta),
\label{2Dsuperfield2}
\end{eqnarray}
where $\theta$-linear functions correspond to 1-forms on $\Sigma$ (we identify $A_1^{+i}$ with $-b^i$),
\beq \uA_i (\sigma, \theta)\equiv \theta^{\mu} A_{\mu i} (\sigma)  \quad \mathrm{and} \qquad 
\uAplus^i (\sigma, \theta) \equiv \theta^{\mu} A_{\mu}^{+i} (\sigma),
\eeq 
and those quadratic in $\theta$ to volume forms,
\beq \uXplus\!\!\!\!_i \,(\sigma, \theta)\equiv  \tfrac{1}{2} \theta^{\mu} \theta^{\nu} \xzero_{\mu\nu i}^{+} (\sigma) \quad \mathrm{and} \qquad 
 \ucplus^i (\sigma, \theta)\equiv \tfrac{1}{2} \theta^{\mu} \theta^{\nu} c_{\mu\nu}^{+i} (\sigma).
\eeq 

The BV symplectic form follows essentially from replacing $\tilx^i$ and $\tila_i$ by 
$\bx^i$ and $\ba_i$ in the integrations on the two dimensional supermanifold, see \eqref{omegaBVtotal} (in our case, $\Sigma_1 = S^1$). Thus, by means of \eqref{superBFVsymplecticform}, we obtain 
\begin{eqnarray} 
\omega_{BV}^{(FHGD)} 
&=& \int_{T[1]\Sigma} d^2 \sigma d^2 \theta \,
\left[ \delta \bx^{i} \delta \ba_i
+ \tfrac{1}{2} 
H_{ijk}(\bx) \bbd_1 \bx^i \delta \bx^j \delta \bx^k
\right].
\label{FHGDBVsymplectic}
\end{eqnarray}
These two terms contain several ones when expressed by means of ordinary fields: 
\begin{eqnarray}
\omega_{BV}^{(FHGD)} 
&=& \int_{T[1]\Sigma} d^2 \sigma d^2 \theta
\left[ \delta X^i \delta \uXplus\!\!\!\!_i - \delta \uA_i \delta \uAplus^i - 
\delta \uc_i \delta \ucplus^i
- H_{ijk}(X) \bbd_1 X^i \delta X^j \delta \uAplus^k \right.
\nonumber \\ &&
\left.- \tfrac{1}{2} H_{ijk}(X) \bbd_1 \uAplus^i  \delta X^j \delta X^k + \tfrac{1}{2}
\partial_l H_{ijk}(X) \uAplus^l \bbd_1 X^i \delta X^j \delta X^k
\right].
\end{eqnarray}


The FHGD-BV action is constructed by substituting 
\eqref{superBFVBRSTcharge} and \eqref{FHGDBVsymplectic}
into \eqref{FHGDactionWZsuper},
\begin{eqnarray}
\Sbvbbd 
&=& \int_{T[1]\Sigma}
d^2 \sigma d^2 \theta \ \left[
\ba_i \bbd \bx^i 
+  \tfrac{1}{2} \pi^{ij}(\bx) \ba_i \ba_j
+ \tfrac{1}{2} \pi^{il} \pi^{jm} H_{klm}(\bx)
\ba_i \ba_j {\bm \varepsilon}_1 \bx^{k} 
\right]
\nonumber \\ &&
+ \frac{1}{3!} \int_{T[1]N} H_{ijk}(\bx) \bbd \bx^i \bbd \bx^j \bbd \bx^k,
\label{superFHGDBVaction}
\end{eqnarray}
where $\bbd \equiv \theta^{\mu} \partial_{\mu}$; here the first term has been obtained by combining originally two terms, one containing $\bbd_0 = \theta^0 \partial_0$ and another one containing $\bbd_1 = \theta^1 \partial_1$, as follows
\begin{eqnarray}
\int_{T[1]\Sigma} d^2 \sigma d^2 \theta \left[
\ba_i \bbd_0 \bx^i + \ba_i \bbd_1 \bx^i 
\right]
&=& 
\int_{T[1]\Sigma}
d^2 \sigma d^2 \theta \  \ba_i \bbd \bx^i . \label{combination}
\end{eqnarray}
The WZ term appears from the $H$-term in the BFV symplectic form, cf.\ the first line in \eqref{FHGDactionWZsuper}. 

By construction, \eqref{superFHGDBVaction} satisfies the classical master equation $$(\Sbvbbd, \Sbvbbd)=0, $$ where the BV bracket $(-,-)$ is defined from the BV symplectic form
\eqref{FHGDBVsymplectic}. Also, again in part by construction, the degree zero part of $\Sbvbbd$ coincides with the classical action \eqref{classicalactionofHPSM} and its degree one part contains a possible set of generators of the gauge transformations. So, from this perspective, the above formulas provide a possible set of the BV data of the HPSM.

However, neither $\omega_{BV}^{(FHGD)}$ nor $\Sbvbbd$ are covariant. 
Although there was some $G$-covariant recombination of terms, see \eqref{combination}, evidently   the explicit appearance of $\bbd_1$ spoils the covariance  in the expression for $\omega_{BV}^{(FHGD)}$, as does ${\bm \varepsilon}_1 \equiv \theta^1 \partial/\partial \theta^1$, the Euler vector field along the ``spatial direction'',  in the expression for $\Sbvbbd$.

For completeness, we also expand the above BV action into component fields:
\begin{eqnarray}
\Sbvbbd
&=& 
\int_{\Sigma} d^2 \sigma \left(
- A_{0i} \partial_1 \xzero^i + A_{1i} \partial_0 \xzero^i 
- \pi^{ij}(\xzero) A_{0i} A_{1j} \right)
+ \int_{N} H
\nonumber \\ 
&&
+ \int_{\Sigma} d^2 \sigma \left[
- \pi^{ij}(\xzero) \xzero^+_{10 i} c_j
+ A_0^{+i} \partial_1 c_i - A_1^{+i} \partial_0 c_i
\right.
\label{FHGDcomponentBVaction}
 \\ 
&&
\quad \quad + A_0^{+k} \pi^{ij},_k(\xzero) A_{1i} c_j
- A_1^{+k} f_k^{ij} A_{0 i} c_j+ \tfrac{1}{2} f_k^{ij} c_{10}^{+k} c_i c_j 
\left.
+ \tfrac{1}{2} f_k^{ij}{},_n A_0^{+n} A_1^{+k} c_i c_j
\right],\nonumber
\end{eqnarray}
with the abbreviation $f_k^{ij}$ has been defined in \eqref{fijk}.


\subsection{The $G$-covariant BV formulation}
\label{secThe}
There is no a priori obstruction for either the BV functional of the HPSM or its BV symplectic form to be invariant with respect to diffeomorphism of $\Sigma$. Such data have been constructed in  \cite{Ikeda-Strobl2} following a more standard construction scheme. We present the result for later comparison:
 
The BV symplectic form can be put into the following form,
\begin{eqnarray} 
\omega_{BV} &=& \int_{T[1]\Sigma} d^2 \sigma d^2 \theta \,
\delta \bx^i \delta \ba_i
\nonumber \\ 
&=& \int_{T[1]\Sigma} d^2 \sigma d^2 \theta \,
(\delta X^i \, \delta \uXplus\!\!\!\!_i - \delta \uA_i \, \delta \uAplus^{i}
- \delta \uc_i \, \delta \ucplus^{i}),
\label{BVsymplecticDarboux}
\end{eqnarray}
while, at the same time, the BV action takes this form:
\begin{eqnarray}
S_{BV} 
&=& 
\int_{T[1]\Sigma} 
\!\!\!\!
d^2 \sigma d^2 \theta \,
\left[\ba_i \, \sd \bx^i 
+ \tfrac{1}{2} \pi^{ij}(\bx)\, \ba_i \ba_{j} \right]
+ \int_{T[1]N} \!\!\!\! d^3 \sigma d^3 \theta \, H(\bx)
\nonumber \\ && 
+ \int_{T[1]\Sigma} \!\!\!\! d^2 \sigma d^2 \theta \, \left[\tfrac{1}{4} 
(\pi^{il} \pi^{jm} H_{lmk})(\bx) \, \ba_i \ba_j \bm{\varepsilon} \bx^k 
+ \tfrac{1}{2} (\pi^{il} H_{jkl}) (\bx)\, \ba_i (\sd \bx^j) \bm{\varepsilon} \bx^k 
\right]
\nonumber \\ &&
+ \int_{T[1]\Sigma} \!\!\!\! d^2 \sigma d^2 \theta \, \left[ \tfrac{1}{8} (\pi^{im} \pi^{jn} \pi^{pq}  H_{mql}H_{npk})(\bx)\,
\ba_i \ba_j (\bm{\varepsilon} \bx^k) \bm{\varepsilon} \bx^l
\right].
\label{superfieldBVaction}
\end{eqnarray}
Both, $\omega_{BV}$ and $S_{BV}$ are explicitly $G$-invariant. And both are quite different from the formulas found by the FHGD formalism---compare, in particular,  \eqref{superFHGDBVaction} to \eqref{superfieldBVaction}. It is noteworthy to remark that albeit the result of the FHGD formalism for the BV action is not covariant, it is significantly shorter than the covariant expression found above. 
This impression remains when \eqref{superfieldBVaction} is expanded in terms of ordinary fields 
\begin{eqnarray}
S_{BV} 
&=& 
\int_{T[1]\Sigma} 
\!\!\!\!
d^2 \sigma d^2 \theta \,
\left(\uA_i \, \sd \uX^i + \tfrac{1}{2} \pi^{ij}  \uA_i \, \uA_j \right) + \int_{T[1]N} 
\!\!\!\!
d^3 \sigma d^3 \theta \,
\tfrac{1}{3!} H_{ijk}\,  \sd X^i \, \sd X^j \, \sd X^k
\nonumber \\ &&
+\int_{T[1]\Sigma} 
\!\!\!\!
d^2 \sigma d^2 \theta \,
\uAplus^i \left[\sd \uc_i + \pi^{jk},_i \uA_j \, \uc_k 
+ \tfrac{1}{2} \pi^{kl} H_{ijl} (\sd \uX^j - \pi^{jl} \uA_l) 
\, \uc_k
\right]
\nonumber \\ &&
+ \int_{T[1]\Sigma} 
\!\!\!\!
d^2 \sigma d^2 \theta \, \left[\uXplus\!\!\!\!_i\: \:\pi^{ji}   \uc_j +
\ucplus^k\: \tfrac{1}{2}  (\pi^{ij},_k + \pi^{il} \pi^{jm} H_{klm}) \,  \uc_i \, \uc_j\right]
\nonumber \\ &&
+ \int_{T[1]\Sigma} 
\!\!\!\!
d^2 \sigma d^2 \theta \,
\tfrac{1}{4} \left[\pi^{ij},_{nk} + (\pi^{il} \pi^{jm} H_{klm}),_n \right] \uAplus^n \uAplus^k \uc_i \, \uc_j
\nonumber \\ &&
+ \int_{T[1]\Sigma} 
\!\!\!\!
d^2 \sigma d^2 \theta \,
\tfrac{1}{8}   \pi^{ci} \pi^{ja} \pi^{bd} H_{nab} H_{kcd} \, \uAplus^n \uAplus^k \uc_i \, \uc_j
\, ,
\label{supercoordBVaction}
\end{eqnarray}
and compared to \eqref{FHGDcomponentBVaction}. 
Besides being shorter, \eqref{superFHGDBVaction}  is also much easier to obtain. One of the main purposes of the rest of this paper is to show first \emph{why} there can be such a difference and second \emph{how} to obtain  from the simple-to-get Fisch--Henneaux--Grigoriev--Damgaard result, in a second step, the covariant expression above.

\subsection[BV from the Fisch--Henneaux--Grigoriev--Damgaard formalism for the BPSM]{BV from the FHGD formalism for the BPSM}
One may ask oneself if possibly all of the non-covariance found in Section \ref{secBVFHGDHPSM} results entirely from the WZ-term. Such a term gives rise to a cohomologically non-trivial contribution to $\omega_{BFV}$, see \eqref{BFVsymplecticform},
 and leads to an evidently non-covariant contribution to the BV symplectic form within the FHGD formalism. This question arises all the more since one observes that in the case of the PSM, i.e.\ for $H=0$ in \eqref{classicalactionofHPSM}, \emph{all} of the non-covariance in   \eqref{FHGDBVsymplectic} and \eqref{superFHGDBVaction} disappears and the FHGD formalism yields the known covariant expressions on the nose. 
 
This is, however, not the case. To illustrate the covariance problems of the FHGD formalism in its present form also without a WZ term, we consider the BPSM, the PSM twisted by a 2-form $B$ governed by the action functional \eqref{BPSM}. In this case, the BFV symplectic form could be presented as in \eqref{BFVsymplecticformXp}, which now leads to the completely covariant BV symplectic form \eqref{BVsymplecticDarboux}. 

Let us next determine the BV functional obtained in this way, using \eqref{FHGDBVaction}  together with \eqref{SBFVfield} and  \eqref{JB}. Integrating out the odd $\theta^0$-variable, one obtains in the end 
\begin{eqnarray}
\Sbvbbd
&=& 
\int_{\Sigma} \left[A_i \wedge dX^i + \tfrac{1}{2} \pi^{ij}(X) A_i \wedge A_j
+ \tfrac{1}{2} B_{ij}(X) dX^i \wedge dX^j
\right.
\nonumber \\ 
&&
\left.
- \pi^{ij}(\xzero) \xzero^+_{i} c_j - A^{+i} \wedge d c_i 
+ \tfrac{1}{2} f_k^{ij}c_{}^{+k} c_i c_j 
\right]
\nonumber \\ 
&&
+ \int_{\Sigma} d^2 \sigma
\left[ A_0^{+k} \pi^{ij}{}_{,k}(\xzero) A_{1i} c_j
- A_1^{+k} f_k^{ij} A_{0 i} c_j
\right.
\nonumber \\ 
&&
\left.
+ \tfrac{1}{2} f_k^{ij}{},_n
A_0^{+n} A_1^{+k} c_i c_j
- c_i \pi^{ij} B_{jk} \partial_1 A_0^{+k} 
+ c_i \pi^{ij} B_{jk}{},_{l} A_0^{+l} \partial_1 \xzero^k
\right].
\label{FHGDBVaction2prime}
\end{eqnarray}
where \eqref{3.25} has been used. The first line is the (covariant) classical action \eqref{BPSM}. Also the second line is still covariant. All the remaining parts are, however, far from covariant (for generic 2-forms $B$---for vanishing $B$, the case of the untwisted PSM, also these terms combine into a covariant expression and then agree with the AKSZ result, i.e.\ in this special case \eqref{FHGDBVaction2prime} coincides with \eqref{superfieldBVaction} for $H$ put to zero).

\section[Mechanical toy models and a useful lemma]{Mechanical toy models and a useful lemma\protect\footnote{In this section we will use the usual notation for mechanical systems, denoting the Hamiltonian by $H$ and a magnetic field 2-form by $B$. This is not to be confused with the 2-form $B$ and the 3-form $H$ that are added to the Poisson sigma model in Sec.\ \ref{FHGDTPSM} above and Sec.\ \ref{secRestore} below so as to yield their twisted versions (by, at the same time, modifying the compatibility with the bivector field certainly). Note also that in the twisted Poisson sigma model there is no non-zero Hamiltonian, cf.\ \eqref{TPSMhamiltonian}.}}
\label{secToy}
In this section we take recourse to mechanical toy models to analyze the situation. This permits us to single out the two independent factors that give rise to the non-covariance problems found above, the Wess-Zumino term on the one hand and the possible addition of contributions proportional to the constraints in the generators of gauge symmetries in the Hamiltonian action functional on the other hand. To cover more general cases than the one of the twisted Poisson sigma model, we also permit, for example, a non-vanishing Hamiltonian (at least, in parts of what will follow in this section). With an appropriate re-interpretation where summations can also include integrals (de Witt notation) much of these considerations generalizes also to field theories and covers in particular the present situation. But we will not make this explicit here, for the sake of simplicity of the notation. 

We first consider the general case of a mechanical model, viewed as a $(0+1)$-dimensional field theory, with first class constraints and compatible Hamiltonian. We determine its gauge symmetries on the level of the associated Hamiltonian functional, which we can treat as the action of the theory (in its first order form). We then specialize to two different scenarios, each of which will be governed by Lie algebroid type of gauge symmetries. We will determine its respective BFV formulation and from it the corresponding BV functional as obtained by the FHGD formalism. In each of the two cases, we contrast it then with a BV functional as it can be constructed directly for the given action functional and establish agreement with the FHGD formalism by appropriate additional transformations. 

In this context we also prove a lemma which states that whenever one has a BV formulation of a gauge theory for one choice of generators of gauge transformations, every other choice can be arranged for by an appropriate BV-canonical transformation. This will be used then to change the Hamiltonian generators of the action functional to gauge generators that include contributions vanishing on the constraint surface, i.e. which are proportional to the constraints. As we will see in the subsequent section, this is one of the main factors needed for covariance of the action functional in the twisted PSM. 

\subsection{The general setting}
Consider a mechanical model with constraints $G_a(x, p)$
and Hamiltonian $H(x, p)$ where the symplectic potential is twisted by a 1-form $A\equiv A_i(x) \rd x^i$. Its associated classical action takes the following form:
\begin{eqnarray}
S_{cl} &=&
\int_{\bR} dt
\, [(p_i - A_i(x))\dot{x}^i - H - \lambda^a G_a],
\label{consraintaction2}
\end{eqnarray}
where $\lambda^a$ are independent variables serving as Lagrange multipliers. 

Evidently, $\pprime_i = p_i - A_i(x)$ is the variable canonically conjugate  to $x^i$; correspondingly, 
$(x^i, p_i)$ have the following non-vanishing Poisson brackets
\begin{eqnarray}
&& \{x^i, p_j\} = \delta^i_j, \label{xp}
\\
&& \{p_i, p_j\} = - \partial_i A_j + \partial_j A_i = - B_{ij}, \label{pp}
\end{eqnarray}
where $B=\rd A$ is a 2-form on the configuration space. In some mechanical applications, it has the interpretation of a magnetic field, with $A$ being the magnetic potential.  The symplectic form corresponding to these brackets has the following form:
\begin{eqnarray}
\omega = \rd x^i \wedge \rd p_i + \tfrac{1}{2} B_{ij}(x) \rd x^i \wedge \rd x^j. \label{canB}
\end{eqnarray} 
This 2-form is symplectic also if $B$ is not exact, but just closed, which can happen if the configuration space $M$ has non-trivial second homology. In that case the action functional \eqref{consraintaction2} does not make sense globally anymore and one needs a Wess-Zumino term, at the expense of compactifying time to a circle: 
\begin{eqnarray}
S_{cl} &=&
\int_{S^1} dt \, 
[p_i \dot{x}^i - H - \lambda^a G_a] + \int_{D} B,
\label{consraintaction1}
\end{eqnarray}
where $\partial D = S^1$ ($D$ could be, for example, a disc) and it is understood that $B$ is pulled back to $D$ in the integral above (with respect to the map $\tilde x \colon D \to M$ where $\tilde x \vert_{ \partial D} = x$).

We assume that the constraints $G_a$ are regular, irreducible, and of first class:
\begin{eqnarray}
\{G_a, G_b \} &=& C_{ab}^c G_c, \label{GG}
\end{eqnarray}
for some functions $C_{ab}^c$. 
Geometrically this means that the constraint surface $G_a=0$ is coisotropic in $T^*M$. We furthermore assume that there are no secondary constraints, i.e.\ the constraints $G_a$ are compatible with the Hamiltonian $H$ of the theory: 
\begin{eqnarray}
\{G_a, H \} &=& V_a^b G_b, \label{GH}
\end{eqnarray}
for some functions $V_a^b$. Geometrically this means that the time evolution generated by the Hamiltonian flow of $H$ does not leave the coisotropic submanifold $G_a =0$. 

We now will derive some consistency conditions following from \eqref{GG} and \eqref{GH}, which we will use later on. First, the Jacobi identity for the Poisson bracket applied to $G_a$, $G_b$, and $G_c$ leads to 
\begin{eqnarray}
&& (C_{[ab}^e C_{c]e}^d + \{G_{[a}, C_{bc]}^d \}) G_d = 0.
\end{eqnarray}
This is equivalent to the existence of functions $\mu_1{}_{abc}^{de}$, completely antisymmetric separately in all upper and lower indices, such that 
\begin{eqnarray}
&& 
C_{[ab}^e C_{c]e}^d + \{G_{[a}, C_{bc]}^d \}  = \mu_1{}_{abc}^{de}\, G_e .
\label{Jacobi1}
\end{eqnarray}
Here we used the following fact: Whenever one has an equation of the sort 
\beq \mu^a G_a = 0, \label{muG}
\eeq 
for some functions $\mu^a$ on $T^*M$ then, necessarily, 
\beq \mu^a \approx 0 , \label{mu}
\eeq 
where the last equality is understood as being valid ``on-shell'', i.e.\ upon restriction to the coisotropic submanifold 
\beq C := \{ (x,p) \in T^*M \vert G_a(x,p)=0 \} .\label{C}\eeq 
 This is the case since by regularity and irreducibility, we can use $G_a$ locally as part of a set of coordinates on phase space and then \eqref{mu} follows from evaluating the derivative of \eqref{muG} with respect  to one of these coordinates at their common zero. It is shown in \cite{Henneaux-Teitelboim}, on the other hand, that \eqref{muG}  and \eqref{mu}  imply the existence of an antisymmetric matrix $\mu^{ab}$ such that 
\beq \mu^a= \mu^{ab}  G_b .
\eeq 
In a similar way, we can conclude from the Jacobi identity for two functions $G$ and the Hamiltonian $H$ that there are (completely antisymmetric) functions $\mu_2{}_{ab}^{cd}$ such that 
\begin{eqnarray}
&& \{G_a, V_b^c \} - \{G_b, V_a^c \} + \{H, C_{ab}^c \} 
- V_a^d C_{bd}^c + V_b^d C_{ad}^c + V_d^c C_{ab}^d = \mu_2{}_{ab}^{cd} G_d.
\label{Jacobi2}
\end{eqnarray}

In what follows, it will be  important to compare the Hamiltonian analysis above with the Lagrangian treatment of the gauge symmetries of the action functional \eqref{consraintaction2} (or \eqref{consraintaction1}). We know that the first class constraints $G_a$ generate the gauge symmetries on the canonical variables $x$ and $p$, albeit only onshell. In this way we arrive at 
\begin{eqnarray}
\delta p_i 
&=& 
- \epsilon^a \frac{\partial G_a}{\partial x^i}
- \epsilon^a B_{ij} \frac{\partial G_a}{\partial p_j}
+ \epsilon^a \omega_{ia}^b G_b,
\label{gaugep}
\\
\delta x^i &=& 
\epsilon^a \frac{\partial G_a}{\partial p_i}
+ \epsilon^a \tau^{ib}_a G_b.
\label{gaugeq}
\end{eqnarray}
Here $\epsilon^a$ are the infinitesimal parameters, which we assume to depend on $t$ only. The first two terms in \eqref{gaugep} and the first term in \eqref{gaugeq} are generated by the Hamiltonian vector field of $\epsilon^a G_a$. The last terms, in which  $\omega_{ia}^b$ and $\tau_a^{ib}$ are arbitrary functions of $t$, $x$, and $p$, are often forgotten, since they parametrize trivial gauge symmetries only (i.e.\ gauge symmetries vanishing on-shell). However, in the application we consider, they will turn out to be essential for the $G$-equivariance. 

The gauge invariance of $S_{cl}$ then fixes the transformation property for $\lambda^a$, except for an arbitrary antisymmetric matrix-valued function $\mu_3^{ab}$:
\begin{eqnarray}
\delta \lambda^a &=& \dot{\epsilon}^a + \epsilon^b \omega_{ib}^a \dot{x}^i
- \epsilon^b \tau_b^{ia} \dot{p}_i - \lambda^b \epsilon^c \left(C_{bc}^a
+ \frac{\partial G_b}{\partial x^i} \tau_c^{ia}
+ \frac{\partial G_b}{\partial p_i} \omega_{ic}^a \right)
\nonumber \\ &&
+ \,\epsilon^b \left(V_{b}^a
+ \frac{\partial H}{\partial x^i} \tau_b^{ia}
+ \frac{\partial H}{\partial p_i} \omega_{ib}^a \right)
+ \mu_3^{ab}G_b.
\label{gaugelambda}
\end{eqnarray}

For the construction of the BV extension of $S_{cl}$ by standard methods, one also needs its Euler-Lagrange equations, which we thus display for convenience. 
\begin{eqnarray}
 \frac{\delta S_{cl}}{\delta p_i} &\equiv&  \dot{x}^i - \frac{\partial H}{\partial p_i} 
- \lambda^a \frac{\partial G_a}{\partial p_i} =0, \label{eom1}
\\
- \frac{\delta S_{cl}}{\delta x^i} &\equiv & \dot{p}_i + B_{ij} \dot{x}^j
+ \frac{\partial H}{\partial x^i} 
+ \lambda^a \frac{\partial G_a}{\partial x^i} =0,\label{eom2}
\\
 \frac{\delta S_{cl}}{\delta \lambda^a} &\equiv & G_a =0.\label{eom3}
\end{eqnarray}

\subsection{Simplifying assumptions and Lie algebroid symmetries}
\label{secSimplifying}

We first want to re-obtain in the present language the statement of \cite{CF-coisotropic} that there is a natural Lie algebroid defined over the constraint surface \eqref{C}. Indeed, introduce a vector bundle $E \to C$ of the rank $r$ that equals the number of (independent) constraints $G_a$. If one assumes that the constraints exist globally, then the bundle can be chosen to be trivial, which we will assume here (one always has $E = N^*C$, the conormal bundle over $C$; here it is important that the codimension of $C$ inside $M$ is also $r$). Let $e_a$ denote the frame spanning $E$ and define the Lie bracket between basis elements by means of the formula
\beq [e_a,e_b]
 := C_{ab}^c \, e_c \, .\label{ee}\eeq 
The functions $C_{ab}^c= C_{ab}^c(x,p)$ were introduced in \eqref{GG}, but now are understood as being restricted to $C$ (i.e.\ pulled back by the embedding map $\iota \colon C \hookrightarrow T^*M$). Similarly, define the anchor by means of the Hamiltonian vector field of $G_a$,
\beq \rho_a \equiv \rho(e_a) := \{ G_a , \cdot \},  \label{rho}
\eeq 
again restricted to $C$; note that here the first class property \eqref{GG} ensures that the Hamiltonian vector fields $\{ G_a , \cdot \}$ are tangential to $C$ and thus \eqref{rho} indeed defines an element in $TC$. Evaluating \eqref{Jacobi1} at $C$ now leads to
\begin{eqnarray}
&& 
C_{[ab}^e C_{c]e}^d + \rho_{[a}(C_{bc]}^d)   \approx 0 ,
\label{JacobiLAapprox}
\end{eqnarray}
which is nothing but the Jacobi identity needed for the bracket \eqref{ee}, extended by means of the Leibniz rule using the map $\rho\colon E \to TC$, to become a Lie bracket and thus define a Lie algebroid on $E\to C$.  

In the rest of this section we will make some simplifying assumptions which will be sufficient for the further analysis, but which constitute a restriction to the general setting. 

\noindent \emph{Assumption 1:}  We require that the function $\mu_1$ in \eqref{Jacobi1} vanishes,
\beq \mu_1{}^{de}_{abc} \equiv 0. \label{mu1}\eeq 
This implies that the above formulas \eqref{ee} and \eqref{rho} define a Lie algebroid 
\beq E \to M \, , 
\eeq 
since, in particular, the decisive equation 
\begin{eqnarray}
&& 
C_{[ab}^e C_{c]e}^d + \rho_{[a}(C_{bc]}^d)   = 0 ,
\label{JacobiLA}
\end{eqnarray}
now holds true on all of $M$. We thus have an off-shell Lie algebroid governing the gauge symmetries of the mechanical models considered in this section. This is something that also holds true for the twisted PSM, see \cite{Severa-Weinstein,Ikeda-Strobl2}, but which, in general, relies only on the single assumption \eqref{mu1} above. This is the main assumption, all the remaining ones are to simplify the (in part heavy) ensuing calculations, while relaxing at least most of them will lead to similar albeit more involved formulas. 

\noindent \emph{Assumption 2:} We restrict our attention to cases where  $C_{ab}^c = C_{ab}^c(x)$ is a function of $x$ only.

This is satisfied also in the twisted PSM (with the obvious replacement of $T^*M$ by $T^*LM$, the cotangent bundle over loops in $M$. 
 
 \noindent \emph{Assumption 3:} We consider the generators for gauge symmetries above only with 
 \beq \tau^{ia}_b:=0 \quad \mathrm{and} \quad \mu_3^{ab} := 0 \, . \label{taumu3}
 \eeq 
 
 \noindent \emph{Assumption 4:} For what concerns the dynamics of the theory, governed by the Hamiltonian $H$, we assume that $H$ is at most quadratic in the momenta and require in addition 
 \beq \mu_2{}_{ab}^{cd} \equiv 0 \quad \mathrm{and} \quad  \{C_{[ab}^{[d}, V_{c]}^{e]}\} \equiv 0.
\label{Identity23}
 \eeq
This fourth assumption may be more restrictive than the second and the third ones and lead to qualitatively different extensions, already in the context of the BFV Hamiltonian. To understand the problem of Sec. \ref{FHGDTPSM}, we do not need a Hamiltonian. However, as one observes from \eqref{gaugelambda}, the gauge symmetries are effected by the Hamiltonian $H$ and in more general, non-topological field theories, restoring covariance in the FHGD method will require taking such terms into account---which we thus do in the treatment of our second toy model. 

 \noindent \emph{Additional assumptions imposed for the first toy model:} In our first model, toy model 1, we want to see how to handle  covariance problems created by Wess-Zumino terms. For this purpose we turn off all dynamics,
 \beq H:= 0 \quad \mathrm{and} \quad V_a^b := 0,
 \eeq 
 and we also eliminate the remaining on-shell vanishing term in the gauge symmetry generators  \eqref{gaugep}, \eqref{gaugeq}, and \eqref{gaugelambda} by putting to zero also the coefficients $\omega$,
 \beq \omega_{ia}^b :=0 \, . 
 \eeq 
 
 \noindent \emph{Additional assumptions imposed for the second toy model:} We showed at the end of the previous section that the covariance problem remains also if the PSM is twisted by a 2-form only, i.e.\ in the absence of WZ terms. Learning from toy model 1 how to deal with WZ terms in general within the FHGD formalism, we will put 
 \beq B := 0,  \eeq 
 in \eqref{consraintaction1}, or, likewise, consider \eqref{consraintaction2} with $A=0$. In addition, we will assume that $\omega_{ia}^b$ depends on $x$ only.


\subsection{Mechanical toy model 1: Wess-Zumino terms}
\label{sec:mechanicsiWZ}
For convenience of the reader we specify the action and gauge symmetries governing the toy model here explicitly. The gauge symmetry of the action
\begin{eqnarray}
S_{cl} &=&
\int_{S^1} dt \, 
[p_i \dot{x}^i  - \lambda^a G_a] + \int_{D} B,
\label{SB}
\end{eqnarray}
are generated as follows:
\begin{eqnarray}
\delta x^i 
&=& \epsilon^a \frac{\partial G_a}{\partial p_i},
\label{gaugeq7} \\
\delta p_i &=& 
- \epsilon^a \frac{\partial G_a}{\partial x^i}
- \epsilon^a B_{ij} \frac{\partial G_a}{\partial p_j},
\label{gaugep7}
\\
\delta \lambda^a &=& \dot{\epsilon}^a - C_{bc}^a \lambda^b \epsilon^c.
\label{gaugelambda7}
\end{eqnarray}

\subsubsection{BV from BFV by means of the FHGD formalism}\label{BFVWZ}
To apply the FHGD formalism, we first need to construct the BFV formulation of the model. For this purpose, we 
introduce a pair of odd coordinates $(c^a, b_a)$ satisfying
\begin{eqnarray}
\{c^a, b_b \} = \delta^a_b,
\end{eqnarray}
so that the total BFV symplectic form becomes
\begin{eqnarray}
\omega_{BFV} = d x^i \wedge d p_i + \frac{1}{2} B_{ij} dx^i \wedge dx^j + d c^a \wedge d b_a.
\label{BFVmechanics1}
\end{eqnarray}
The BFV-BRST charge is then simply 
\begin{eqnarray}
S_{BFV} &=& c^a G_a - \frac{1}{2} C_{ab}^c b_c c^a c^b.
\end{eqnarray}
It satisfies $\{S_{BFV}, S_{BFV}\}=0$, which one shows by 
using the Lie algebroid identity \eqref{JacobiLA} (together with \eqref{rho} and \eqref{GG}).

Now we are in the position to apply the formalism explained in Sec.\ \ref{secFHGD}. We thus introduce the following superfields
\begin{eqnarray}
X^i &=& x^i - \theta^0 p^{*i},
\\
P_i &=& p_i + \theta^0 x^*_i,
\\
C^a &=& - c^a + \theta^0 b^{*a} ,
\\
B_a &=& b_a + \theta^0 c^*_a .
\label{FHGDsuperfield}
\end{eqnarray}
For later comparison we put 
\beq \lambda^a \equiv  b^{*a} \qquad \mathrm{and} \quad \lambda^*_a \equiv b_a. \label{Notation} \eeq
The FHGD-BV symplectic form results from the BFV symplectic form \eqref{BFVmechanics1},
\begin{eqnarray}
\omega_{BV}^{FHGD} 
&=& \int_{T[1]S^1} d \theta^0 d t (\delta X^i \wedge \delta P_i 
+ \tfrac{1}{2} B_{ij}(X) \delta X^i \wedge \delta X^j + \delta C^a \wedge \delta B_a),
\end{eqnarray}
which, written out in component fields, takes the form
\begin{eqnarray}
\omega_{BV}^{FHGD} 
&=& \int_{S^1} dt\left[\delta x^i \wedge \delta x^*_i + \delta p_i \wedge \delta p^{*i} 
+ \delta c^a \wedge \delta c^*_a + \delta \lambda^a \wedge \delta \lambda^*_a \right.
\nonumber \\ &&
\left. \qquad \: \, + \, B_{ij}(x) \delta x^i \wedge \delta p^{*j} 
+ \tfrac{1}{2} \partial_k B_{ij}(x) p^{*k} \delta x^i \wedge \delta x^{j} \right].
\label{FHGDBVsymplecticWZ}
\end{eqnarray}
The BFV symplectic form is not exact. So for the construction of the BV functional we need to use \eqref{FHGDWZ}, developed in the context of WZ terms. This yields
\begin{eqnarray}
S_{BV}^{FHGD} 
&=& \int_{T[1]S^1} d \theta^0 d t (P_i \rd X^i  - B_a \bbd_0 C^a) + \int_D B(x)
\nonumber \\ &&
- \int_{T[1]S^1} d \theta^0 dt \left(C^a G_a(X, P) - \frac{1}{2} C_{bc}^a(X) B_a C^b C^c
\right).
\end{eqnarray}
Using $\bbd_0 \equiv \theta^0 \frac{d}{dt}$ and performing the odd integration over $\theta^0$ we obtain
\begin{eqnarray}
S_{BV}^{FHGD} 
&=& \int_{S^1} d t [p_i\dot{x}^i + b^a \dot{c}_a - b^{*a} G_a] + \int_{D} B
\nonumber \\ &&
+ \int_{S^1} \left[x^*_i \frac{\partial G_a}{\partial p_i} c^a 
- p^{*i} \left( \frac{\partial G_a}{\partial x^i} c^a + \frac{1}{2} \frac{\partial C_{bc}^a}{\partial x^i} b_a c^b c^c \right)
- \frac{1}{2} C_{bc}^a c^*_a c^b c^c - C_{bc}^a b_a b^{*b} c^c \right]
\nonumber \\
&=& \int_{S^1} d t [p_i \dot{x}^i - \lambda^{a} G_a]
+ \int_{D} B + \int_{S^1} \left[x^*_i \frac{\partial G_a}{\partial p_i} c^a - p^{*i} \frac{\partial G_a}{\partial x^i} c^a 
\right.
\nonumber \\ &&
\left.
+ \lambda^{*a} (\dot{c}_a - C_{bc}^a \lambda^b c^c)
- \frac{1}{2} C_{bc}^a c^*_a c^b c^c 
+ \frac{1}{2} \frac{\partial C_{bc}^a}{\partial x^i} \lambda^*_a p^{*i} c^b c^c \right],
\label{FHGDBVactionWZ}
\end{eqnarray}
where we used \eqref{Notation}.

\subsubsection{BV formalism from standard methods}\label{BVWZ}
Here we construct a BV extension $S_{BV}$ using the normal procedure. Replacing $\epsilon^a$ by an  odd ghost $c^a$ in the gauge transformations, we obtain for the BRST operator $s$: 
\begin{eqnarray}
s x^i &=& 
\frac{\partial G_a}{\partial p_i} c^a ,
\label{BRSTq}
\\
s p_i &=& - \frac{\partial G_a}{\partial x^i} c^a
- B_{ij} \frac{\partial G_a}{\partial p_j} c^a,
\label{BRSTp}
\\
s \lambda^a &=& \dot{c}^a - C_{bc}^a \lambda^b c^c. 
\label{BRSTlambda}
\end{eqnarray}
For the action of the BRST operator on the ghosts, we use the structure functions appearing in \eqref{GG}:\footnote{We may consider more general expressions for $s c^a$ by adding terms vanishing on-shell, but we will not pursue this here.}
\begin{eqnarray}
s c^a = \frac{1}{2} C_{bc}^a c^b c^c.
\end{eqnarray}
The BRST transformations need to square to zero at least on-shell, i.e., in this context, upon the usage of the Euler-Lagrange equations \eqref{eom1}, \eqref{eom2}, and \eqref{eom3} (here with $H=0$). This is indeed the case: 
\begin{eqnarray}
s^2 x^i &=& 0,\\
s^2 p_i &=& \frac{1}{2} \frac{\partial C_{bc}^a}{\partial x^i} G_a c^b c^c
\equiv - \frac{1}{2} \frac{\partial C_{bc}^a}{\partial x^i} \frac{\delta S_{cl}}{\delta \lambda^a} c^b c^c,
\\
s^2 \lambda^a &=& 
\frac{1}{2} \frac{\partial C^a_{bc}}{\partial x^i} \left(\dot{x}^i - \frac{\partial G_d}{\partial p_i} \lambda^d \right) c^b c^c
\equiv \frac{1}{2} \frac{\partial C^a_{bc}}{\partial x^i} \frac{\delta S_{cl}}{\delta p_i} c^b c^c.
\end{eqnarray}
Although, in general, the need for the BV method arises when one has an ``open algebra'' (i.e.\ when the symmetries of the action functional cannot be written as a semi-direct product of trivial and the on-shell non-vanishing ones), general experience shows that this happens when one has field-dependent structure functions in the constraint algebra. This is also confirmed in the present toy model: only when one has structure constant $C^a_{bc}$, corresponding to a finite-dimensional Lie algebra, the BRST operator squares to zero already off-shell. 

The BV symplectic form is the canonical symplectic form with 
conjugate pairs for each field and its antifield:
\begin{eqnarray}
\omega_{BV} 
&=& \int_{\bR} dt(\delta x^i \wedge \delta x^*_i + \delta p_i \wedge \delta p^{*i} + \delta c^a \wedge \delta c^*_a + \delta \lambda^a \wedge \delta \lambda^*_a).
\label{BVsymplecticWZ}
\end{eqnarray}
The BV action $S_{BV}$ is computed as follows.
We determine $S_{BV}$ by expanding it by the order of antifields as
\begin{eqnarray}
S_{BV} 
&=& S_{BV}^{(0)} + S_{BV}^{(1)} + S_{BV}^{(2)} + \cdots,
\end{eqnarray}
where $S_{BV}^{(0)}$ is the classical action, 
$S_{BV}^{(1)} = (-1)^{|\Phi|} \int \Phi^* s \Phi$ is determined from gauge transformations and  $S_{BV}^{(2)}$ is determined from $s^2 \Phi$.
\begin{eqnarray}
S_{BV}^{(0)} &=& S_{cl} = \int_{} dt [p_i \dot{x}^i - \lambda^a G_a]
+ \int_{\Sigma} B,
\\
S_{BV}^{(1)} &=& \int_{} dt [x^*_i s x^i + p^{*i} s p_i + \lambda^*_a s \lambda^a - c^{*a} s c_a]
\nonumber \\ 
&=& \int_{} dt \left[
x^*_i \frac{\partial G_a}{\partial p_i} c^a 
+ p^{*i} \left(- \frac{\partial G_a}{\partial x^i}
- B_{ij} \frac{\partial G_a}{\partial p_j} \right) c^a 
\right.
\nonumber \\ && 
\left.
+ \lambda^*_a (\dot{c}^a - C_{bc}^a \lambda^b c^c) - c^{*a} \frac{1}{2} C_{bc}^a c^b c^c
\right],
\\
S_{BV}^{(2)} &=& \int_{} dt \left(\frac{1}{4} \frac{\partial C_{bc}^a}{\partial x^i} \lambda^*_a p^{*i} c^b c^c  
- \frac{1}{4} \frac{\partial C_{bc}^a}{\partial x^i} p^{*i} \lambda^*_a c^b c^c \right)
\nonumber \\ &=& 
\int_{} dt \frac{1}{2} \frac{\partial C_{bc}^a}{\partial x^i} \lambda^*_a p^{*i} c^b c^c.
\end{eqnarray}
The higher terms of $S_{BV}$, $S_{BV}^{(I)}$ for $I=3,4,\ldots$, need to be chosen such that the classical master equation $(S_{BV}, S_{BV})=0$ is satisfied, where $(-,-)$ is the BV bracket induced from the BV symplectic form \eqref{BVsymplecticWZ}.
It turns out that we do not need any higher contributions here. Thus, in total we obtain the following BV extension of \eqref{SB}
\begin{eqnarray}
S_{BV} &=& \int_{S^1} dt [p_i \dot{x}^i - \lambda^a G_a]
+ \int_{D} B
+ \int_{S^1} dt \left[
x^*_i \frac{\partial G_a}{\partial p_i} c^a 
+ p^{*i} \left(- \frac{\partial G_a}{\partial x^i}
- B_{ij} \frac{\partial G_a}{\partial p_j} \right) c^a 
\right.
\nonumber \\ && 
\left.
+ \lambda^*_a (\dot{c}^a - C_{bc}^a \lambda^b c^c) - \frac{1}{2} C_{bc}^a c^{*a} c^b c^c
+ \frac{1}{2} \frac{\partial C_{bc}^a}{\partial x^i} \lambda^*_a p^{*i} c^b c^c \right].
\label{BVactionWZ}
\end{eqnarray}

\subsubsection{Comparison and non-canonical change of variables}
Comparing the two results \eqref{FHGDBVactionWZ} and \eqref{BVactionWZ}  to one another, we first observe that they agree for $B=0$. This is in part due to our simplifying assumptions. As we noticed already for the 2-form twisted Poisson sigma model in \ref{BPSM} and we will see also in 
Sec.\ \ref{sec:mechanicstirivalgauge}, where we will put the WZ-term to zero, $B:=0$, there still can be a problem with an agreement of the two approaches---which, at least  in some field theories, turns out to be decisive in the context of spacetime covariance. And it is important to separate these two effects. 

The issue with the WZ term is already evident in the BV symplectic form that one obtains from the FHGD formalism. Whenever one has a WZ term in the theory, $\omega_{BV}^{FHGD}$ does no more agree with the standard BV symplectic form $\omega_{BV}$, which, by construction with the usual procedure (see, e.g., \cite{Henneaux-Teitelboim}), always comes in Darboux coordinates, with the antifields being the momenta of the classical fields and the ghosts. In the present mechanical situation, this discrepancy is visible by comparing \eqref{FHGDBVsymplecticWZ} to \eqref{BVsymplecticWZ}. 

In this toy model it is now easily verified that there is a simple transformation of fields which, at the same time, maps \eqref{FHGDBVsymplecticWZ} to \eqref{BVsymplecticWZ} and \eqref{FHGDBVactionWZ} to \eqref{BVactionWZ}. This transformation changes only the antifield of $x$:
\beq x^*_i \mapsto x^*_i + B_{ij}(x) p^{*j}. \label{WZchange}
\eeq 

In general, there is always a transformation that brings $\omega_{BV}^{FHGD}$ to $\omega_{BV}$. However, not always will this lead to a simultaneous agreement on the level of the BV functionals. This will be particularly obvious in Sec.\ \ref{sec:mechanicstirivalgauge}, where $\omega_{BV}^{FHGD}$ agrees already with $\omega_{BV}$ to start with, but there are also potential pitfalls also in the current, simpler setting, which we consider worth being pointed out.

Suppose, for a moment, that $B$ is in fact exact, $B= \rd A$.  Recall that in such a situation $p'_i = p_i - A_i(x)$ is the momentum canonically conjugate to $x^i$. Let us perform the corresponding change of variables on the level of the superfields on $T[1] S^1$ (or $T[1] \R$, if one prefers):\begin{eqnarray}
P_i^{\prime} = P_i - A_i(X).
\end{eqnarray}
This is equivalent to
\begin{eqnarray}
\pprime_i = p_i - A_i(x), \label{change1}
\\
\xprime^*_i = x^*_i - \partial_j A_i(x) p^{*j}.\label{change2}
\end{eqnarray}
It is now not difficult to see, and also not so surprising, that this transformation turns the BV symplectic form \eqref{FHGDBVsymplecticWZ} to the canonical BV form \eqref{BVsymplecticWZ} (by identifying the primed fields in the first expression with the unprimed ones in the second one).
However, implementing the change of variables \eqref{change1} and \eqref{change2} in \eqref{FHGDBVactionWZ}, one finds 
\begin{eqnarray}
S_{FHGD}^{\prime}
&=& \int_{\bR} d t [\pprime_i \dot{x}^i + \lambda^{*a} \dot{c}_a - \lambda^{a} G_a(x, \pprime + A)
\nonumber \\ &&
+ x^{*\prime}_i \frac{\partial G_a}{\partial p_i} c^a 
- p^{*i} \left( \frac{\partial G_a}{\partial x^i} 
- \partial_i A_j \frac{\partial G_a}{\partial p_j} \right) c^a 
- \frac{1}{2} C_{bc}^a c^*_a c^b c^c - C_{bc}^a \lambda^*_a \lambda^b c^c
\nonumber \\ &&
+ \frac{1}{2} \frac{\partial C_{bc}^a}{\partial x^i} \lambda^*_a p^{*i} c^b c^c ].
\end{eqnarray}
Now again dropping the primes for an identification with $S_{BV}$, we not only do not find an agreement with \eqref{BVactionWZ}, but even worse: If we set all antifields to zero, we are supposed to find the classical action  \eqref{consraintaction1}. This is no more the case here, $S_{FHGD}^{\prime}|_{cl}\neq S_{cl}$. 

This can be corrected by performing a now canonical (i.e.\ BV-symplectic) transformation, which undoes the transformation on the momenta $p_i$ (but now preserves  $\omega_{BV}$): 
\begin{eqnarray}
p_i^{\prime\prime} &=& \pprime_i + A_i(x),
\label{Mechchange3}
\\ 
x^{*\prime\prime}_i &=& \xprime^*_i + \partial_i A_j(x) p^{*j}.
\label{Mechchange4}
\end{eqnarray}
Indeed, combined transformations, composing  \eqref{change1} and \eqref{change2} with  \eqref{Mechchange3} and \eqref{Mechchange4}, we find
\begin{eqnarray}
p_i^{\prime\prime} &=& p_i,
\label{Mechchange5}
\\ 
x^{*\prime\prime}_i &=& x^*_i + B_{ij}(x) p^{*j}.
\label{Mechchange6}
\end{eqnarray}
This now reproduces \eqref{WZchange} and not only has all the desired properties, but also makes sense for a magnetic field $B$ that is not exact. 


\subsection{Change of symmetry generators as a BV symplectic transformation}
\label{secChange}
The remaining problem of  the FHGD formalism stems from the fact that sometimes spacetime covariance requires the inclusion of particular on-shell vanishing contributions to the gauge symmetries, i.e., in the case of the mechanical model, the terms proportional to $G_a$ in the gauge transformations \eqref{gaugeq}, \eqref{gaugep}, and  \eqref{gaugelambda}. Such terms are missing in the non-improved version of the FHGD formalism. Including them, corresponds to a change of the generators of the (essential) gauge symmetries. In this section we want to show that \emph{every} such a change can be implemented by a canonical (BV-symplectic) transformation. 

Let us recall the general setting of infinitesimal gauge symmetries. They always give rise to the following exact sequence of Lie algebras
\beq  0 \to \mathfrak{g}_{triv} \to  \mathfrak{g} \to  \mathfrak{g}_{ess} \to 0 \, . \label{seq}
\eeq 
Here $\mathfrak{g}_{triv}$ denotes (the Lie algebra of) those gauge symmetries, which vanish when applied to solutions of the Euler-Lagrange equations of the given functional $S$, they are on-shell vanishing, $\mathfrak{g}$ denotes all the gauge symmetries of $S$, and $\mathfrak{g}_{ess}$ is the corresponding quotient Lie algebra, which corresponds to the essential part of the gauge symmetries. If one talks of generators of gauge transformations, one actually means representatives of a basis of elements in $\mathfrak{g}_{ess}$, but written as concrete vector fields which annihilate the action functional and which thus live in $\mathfrak{g}$. In other words, one needs to choose a splitting of \eqref{seq} in terms of vector spaces. Every \emph{change} of such a splitting then corresponds to the addition of on-shell vanishing parts to the previous generators. 

To be more concrete, we will use the condensed DeWitt notation: $x^I$ denote the fundamental fields, assumed to be bosonic here for simplicity, with the index  
$I$ including discrete labels $i$ as well as ``continuous ones'' $\sigma^{\mu}$. A summation over indices $I$ correspond, simultaneously, to a summation over $i$ and an integration over                  
$\sigma^{\mu}$. $S = S(x)$ is the classical action. Let $\epsilon \in \mathfrak{g}_{ess}$ and denote by $\delta_\epsilon$ the corresponding generators of $\mathfrak{g}$ after the choice of a splitting. This means, in particular, that one is given some formulas for $\delta_\epsilon x^I$ such that 
\begin{eqnarray}
\delta_\epsilon  S&=& 0.
\end{eqnarray}
Denote by 
\beq \delta_\epsilon' x^I = \delta_\epsilon x^I + \delta^0_\epsilon x^I,
\eeq  another set of such generators. Then there always exists an ``antisymmetric matrix'' $\Lambda^{IJ}$, $\Lambda^{IJ} = - \Lambda^{JI}$, depending on the fields $x^J$ and linearly on $\epsilon$, such that \cite{Henneaux-Teitelboim}
\begin{eqnarray}
\delta_\epsilon^0 x^I &=& \Lambda^{IJ} \frac{\delta S}{\delta x^J}.
\label{trivialgaugetransformatoin}
\end{eqnarray}
Here $\frac{\delta S}{\delta x^I}$ denotes a functional derivative of $S$. Certainly, by construction, $\delta_\epsilon'
 S=0$, but one verifies this also by the antisymmetry of $\Lambda$:
\begin{eqnarray}
\delta_\epsilon^0 S &=& \frac{\delta S}{\delta x^I} \delta_\epsilon^0 x^I
= \frac{\delta S}{\delta x^I} \Lambda^{IJ} \frac{\delta S}{\delta x^J} =0.
\end{eqnarray}

We now assume that we are given a BV extension \begin{eqnarray}
S_{BV} &=& S^{(0)} + S^{(1)} + S^{(2)} + \ldots, \label{SBVorder}
\end{eqnarray}
of the classical action $S=S^{(0)}$ using the generators $\delta_\epsilon$. This means, in particular, that one introduced antifields $x_I^*$ for $x^I$ and $c^*_a$ for $c^a$, where $c^a$ are the ghosts corresponding to a basis $\epsilon_a$ of $\mathfrak{g}_{ess}$, with the BV symplectic form (of degree minus one) looking as follows
\begin{eqnarray}
\omega_{BV} &=& \delta x^I \wedge \delta x^*_I + \delta c^a \wedge \delta c^*_a.
\end{eqnarray}
The superscript in \eqref{SBVorder} denotes the order in the antifields. The choice of generators 
of the gauge transformations enters the formula for $S_{BV}$ only implicitly, namely it is a part of $S^{(1)}$. More precisely, the in general only on-shell nilpotent BRST operator $s$, when acting on the classical fields, takes the form $s (x^I) = \delta_{\epsilon_a} x^I \, c^a$. In this notation, one has
\begin{eqnarray}
S^{(1)} &=& x^*_I \, s x^I - c^*_a \, s c^a,
\end{eqnarray}
where, by definition, $s c^a$ gives the action of the BRST operator on the ghosts. The choice of terms  $S^{(2)} + \ldots$ in $S_{BV}$ is not unique, but such that $S_{BV}$ satisfies the classical master equation, 
\begin{eqnarray}
~( S_{BV}, S_{BV} ) &=& 0,
\end{eqnarray}
where the BV bracket $(-,-)$ (of degree plus one) corresponds to $\omega_{BV}$.

We now want to construct from these data a possible BV extension of $S$ for the choice of generators $\delta'$, i.e. for the BRST operator 
\begin{eqnarray}
s' x^I = s x^I + s_0 x^I, 
\end{eqnarray}
where $s_0$ is corresponds to the above trivial gauge transformations: 
\begin{eqnarray}
s_0 x^I &=& \Lambda^{IJ}(x, c) \frac{\delta S}{\delta x^J},
\label{trivialBRSTtransformatoin}
\end{eqnarray}
with $\Lambda^{IJ}(x, c) \equiv \Lambda_{a}^{IJ}(x) c^a$.

\begin{lemma}\label{lemmact}
\begin{eqnarray}
S'_{BV} &:=& e^{H_{\Lambda}} S_{BV},
\end{eqnarray} 
is a BV extension of $S$ with BRST operator $s'$ as given above. Here $H_{\Lambda}$ is the BV Hamiltonian vector field for the functional $\Lambda$ where 
\begin{eqnarray}
\Lambda &=& 
\tfrac{1}{2} \Lambda^{IJ}_{a}(x) x^*_I x^*_J c^a. \label{Lambda}
\end{eqnarray}
\end{lemma}
Proof: $\Lambda$ is of degree minus one. Since the BV bracket is of degree plus one, the Hamiltonian vector field $H_\Lambda := (\Lambda , -)$ is of degree zero and thus its exponential is a degree preserving isomorphism of the BV phase space. Such transformations preserve BV brackets (essentially due to the graded Jacobi identity satisfied by the bracket), 
\begin{eqnarray}
(e^{H_{\Lambda}} F_1, e^{H_{\Lambda}} F_2) = e^{H_{\Lambda}} (F_1, F_2),
\label{canoncaltransfBV}
\end{eqnarray}
for all functionals $F_1, F_2$. From this it is clear that $S'_{BV}$ satisfies the master equation, $( S'_{BV}, S'_{BV} ) = 0$. On the other hand, $\Lambda$ is quadratic in the antifields, thus $H_{\Lambda}$ is at least linear. Correspondingly, to order zero, nothing changes, $S^{'(0)} = S^{(0)} = S$. It remains to be checked that the BRST operator changes in the wished-for way. It is now easy to see that 
\begin{eqnarray}
S^{'(1)} &=& 
x^*_I s x^I - c^*_a s c^a  + x^*_I s_0 x^I
= x^*_I s' x^I - c^*_a s' c^a,
\end{eqnarray}
which concludes the proof. 

We finally remark that, written like this, the change from $S_{BV}$ to $S'_{BV}$ looks rather innocent. In practice, however, the difference in the higher order contributions can explode. One gets a first impression of this phenomenon already for the second toy model, discussed below, as well as for the HPSM---and this despite of the fact that, in both cases, the BV functionals are still at most quadratic in the antifields.

\subsection{Mechanical toy model 2: trivial gauge transformations}
\label{sec:mechanicstirivalgauge}
In this section we consider the action functional
\begin{eqnarray}
S &=&
\int_{} dt
(p_i \dot{x}^i - H - \lambda^a G_a ),
\label{consraintaction}
\end{eqnarray}
which gives rise to the canonical symplectic form 
\begin{eqnarray}
\omega = dx^i \wedge dp_i,
\end{eqnarray}
and for its (essential) gauge transformations the following generators
\begin{eqnarray}
\delta_\epsilon x^i &=& \epsilon^a \frac{\partial G_a}{\partial p_i},
\label{gaugeq3}\\
\delta_\epsilon p_i &=& - \epsilon^a \frac{\partial G_a}{\partial x^i}
+ \epsilon^a \omega_{ia}^b G_b,
\label{gaugep3}
\\
\delta_\epsilon \lambda^a &=& \dot{\epsilon}^a - C_{bc}^a \lambda^b \epsilon^c 
+ V_{b}^a \epsilon^b 
+ \omega_{ib}^a \left(\dot{x}^i 
- \lambda^c \frac{\partial G_c}{\partial p_i} 
- \frac{\partial H}{\partial p_i} \right) \epsilon^b.
\label{gaugelambda3}
\end{eqnarray}

\subsubsection{The FHGD functional}\label{BFVtrivial}
The BFV symplectic form and the BFV-BRST charge are readily determined to be
\begin{eqnarray}
\omega_{BFV} = d x^i \wedge d p_i + d c^a \wedge d b_a,
\label{BFVsymplectictrivial}
\end{eqnarray}
and
\begin{eqnarray}
S_{BFV} &=& c^a G_a - \frac{1}{2} C_{ab}^c b_c c^a c^b,
\label{BFVfunctionaltrivial}
\end{eqnarray}
respectively. Under the assumptions specified in Sec.\ \ref{secSimplifying}, the BFV extension of the Hamiltonian $H$ can be taken to be of the form 
\begin{eqnarray}
H_{BFV} &=& H + c^a V_a^b b_b.
\end{eqnarray}
It satisfies \eqref{BFVequation2} and \eqref{BFVequation3}. We now may follow the general steps  reviewed in Sec.\ \ref{traditional} (see also Sec.\ \ref{BFVWZ}, but now including the additional Hamiltonian). This yields 
\begin{eqnarray}
\omega_{BV}^{FHGD} 
&=& \int_{\bR} dt(\delta x^i \wedge \delta x^*_i + \delta p_i \wedge \delta p^{*i} 
+ \delta c^a \wedge \delta c^*_a + \delta \lambda^a \wedge \delta \lambda^*_a),
\label{FHGDBVsymplectictrivial}
\end{eqnarray}
and 
\begin{eqnarray}
S_{BV}^{FHGD} 
&=& \int_{\bR} d t [p_i \dot{x}^i  - H - \lambda^{a} G_a
\nonumber \\ &&
+ x^*_i \frac{\partial G_a}{\partial p_i} c^a - p^{*i} \frac{\partial G_a}{\partial x^i} c^a 
+ \lambda^*_a (\dot{c}^a - C_{bc}^a \lambda^b c^c
+ V^a_b c^b)
- \frac{1}{2} C_{bc}^a c^*_a c^b c^c 
\nonumber \\ &&
+ \frac{1}{2} \frac{\partial C_{bc}^a}{\partial x^i} \lambda^*_a p^{*i} c^b c^c 
].
\label{FHGDBVactiontrivial}
\end{eqnarray}

\subsubsection{Standard BV formalism}\label{BV}
In this section  we may follow again the standard procedure of finding the BV extension. As this goes very much in parallel to what we did in Sec.\ \ref{BVWZ}, we will be much briefer here. We remark, however, that the additional terms in \eqref{gaugep3} and \eqref{gaugelambda3} together with the presence of a non-vanishing Hamiltonian contribution to the classical action $S$ complicate the situation considerably. While the BV symplectic form takes again the standard canonical form \eqref{BVsymplecticWZ}, now in complete agreement also with \eqref{FHGDBVsymplectictrivial}, the BV functional becomes 
\begin{eqnarray}
S_{BV} &=&  \int_{\bR} dt \, [p_i \dot{x}^i - H - \lambda^a G_a] \nonumber 
\\
&&+ \int_{\bR} dt \left[
x^*_i \frac{\partial G_a}{\partial p_i} c^a 
+ p^{*i} \left(- \frac{\partial G_a}{\partial x^i} 
+ \omega_{ia}^b G_b \right) c^a 
\right.
\nonumber \\ && 
\left.
\qquad + \lambda^*_a \left(\dot{c}^a - C_{bc}^a \lambda^b c^c + V^a_b c^b 
+ \omega_{ia}^b \left[\dot{x}^i -  \frac{\partial H}{\partial p_i}
- \lambda^c \frac{\partial G_c}{\partial p_i}\right] c^a\right)
- c^{*a} \frac{1}{2} C_{bc}^a c^b c^c
\right]
\nonumber \\
&& 
+ \int_{\bR} dt 
\left[\left(\frac{1}{2} \frac{\partial C_{bc}^a}{\partial x^i} 
+ \frac{\partial \omega_{ib}^a}{\partial x^i} \frac{\partial G_c}{\partial p_i} - \frac{\partial G_c}{\partial x^i} \frac{\partial \omega_{ib}^a }{\partial p_i}  - \omega_{jb}^a \frac{\partial^2 G_c}{\partial x^i \partial p_j} \right) \lambda^*_a p^{*i} c^b c^c \right.\nonumber \\
&&
\qquad  + \left(C_{db}^a \omega_{ic}^d + \frac{1}{2} C_{bc}^d \omega_{id}^a
+ \frac{1}{2} \omega_{jb}^a \omega_{ic}^d \frac{\partial G_d}{\partial p_j} 
\right) \lambda^*_a p^{*i} c^b c^c 
\nonumber \\ &&
\left. \qquad - \frac{\partial^2 G_b}{\partial p_i \partial p_j} \omega_{jc}^a 
\lambda^*_a x^*_i c^b c^c
+ \left(\frac{\partial V^a_{b}}{\partial p_i} \omega_{ic]}^d 
+ \frac{1}{2} \omega_{ib}^a \omega_{jc}^d 
\frac{\partial^2 H}{\partial p_i \partial p_j} \right)
\lambda^*_a \lambda^*_d c^b c^c\right].
\label{BVactiontrivial}
\end{eqnarray}
Note that in the first toy model there was only one term quadratic in the antifields, see \eqref{BVactionWZ}, and, even more surprisingly, the same is the case for the BV functional that one obtains in the FHGD formalism for this model, see \eqref{FHGDBVactiontrivial}, while here there are two lines of such terms---and the verification that the above expression satisfies indeed $(S_{BV} , S_{BV}) = 0$ is a correspondingly much more intensive calculational task.

\subsubsection{Applying the lemma}
We observed in Sec.\ \ref{secChange} that a first comparison of two BV extensions of the same classical action, as here \eqref{FHGDBVactiontrivial} and \eqref{BVactiontrivial}, one should look at the difference to first order in the antifields, from which one can read off the BRST operator $s$. It is not difficult to see that for the extension  \eqref{BVactiontrivial} on has 
\begin{eqnarray}
s x^i &=& 
\frac{\partial G_a}{\partial p_i} c^a,
\label{BRSTq3}\\
s p_i &=& - \frac{\delta G_a}{\delta x^i} c^a
- \frac{\delta S}{\delta \lambda^a} \omega_{ia}^b c^a,
\label{BRSTp3}
\\
s \lambda^a 
&=& \dot{c}^a - C_{bc}^a \lambda^b c^c
+ V_b^a c^b + \frac{\delta S}{\delta p_i} \omega_{ic}^a c^c,
\label{BRSTlambda3}
\end{eqnarray}
which differs from the BRST operator obtained from \eqref{FHGDBVactiontrivial} precisely by the terms proportional to $\omega_{ia}^b$, all of which are on-shell vanishing. We thus can apply the lemma of the previous subsection so as to guarantee agreement of at least the terms to first order in the antifields. 

Comparison of the above formulas with \eqref{trivialBRSTtransformatoin} and \eqref{Lambda} shows that here the generating functional $\Lambda$ takes the form 
\beq \Lambda = \int_{\bR} dt \, \omega_{ia}^b \, c^a \, \lambda^*_b \, p^{*i}.
\eeq 
An explicit calculation now establishes that this transforms  \eqref{FHGDBVactiontrivial} altogether into 
\eqref{BVactiontrivial}, 
\beq S_{BV} = \exp(H_\Lambda) \cdot S_{BV}^{FHGD} . \label{expH}
\eeq

\section{Restoring covariance for the HPSM}
\label{secRestore}
In this section we will now return to the twisted Poisson sigma model and show how one can arrive from its easy-to-obtain but non-covariant BV-formulation  \eqref{FHGDBVsymplectic} and \eqref{superFHGDBVaction} at the covariant BV formulas of \cite{Ikeda-Strobl2} (recalled in Sec.\ \ref{secThe}). Equipped with the insights from the last section, this will be rather straightforward. 

We start with the BV symplectic form. 
Assuming for a moment that the twisting is by means of a 2-form $B$ (or with an exact $H=\rd B$),  one might think that the easiest way to bring $\omega_{BV}^{(FHGD)}$ into Darboux form would be the simple shift transformation 
\begin{eqnarray}
\ba_i \mapsto \ba_i - B_{ij}(\bx) \bbd_1 \bx^j.
\label{changingofA}
\end{eqnarray}
While this indeed serves the purpose for what concerns the BV 2-form, it leads to the same problem as for the mechanical toy model: it destroys the required property that the antifield-free part $S_{BV}^{(0)}$  of $S_{BV}$ has to agree with the classical action. Indeed, as a calculation shows, the change \eqref{changingofA} leads to 
\begin{align}
\left(\Sbvbbd\right)^{(0)} \mapsto &
\int_{\Sigma} \left(A_i \wedge \rd X^i + \tfrac{1}{2} \pi^{ij}(X) A_i \wedge A_j
\right) + \\
& \int_{\Sigma} d^2 \sigma \left(
- \pi^{ij} B_{jk} (c_i \partial_1 A_0^{+k} - A_{0i} \partial_1 \xzero^k)
\right),
\label{FHGDBVaction2classical}
\end{align}
which does not agree anymore with the classical action \eqref{BPSM} (or \eqref{classicalactionofHPSM} for $H=\rd B$).

The analogue to \eqref{WZchange} for the mechanical model  turns out to be 
\begin{eqnarray}
X_{01i}^{+\prime} &:=& X_{01i}^{+} - H_{ijk}(\xzero) A_0^{+k} \partial_1 \xzero^j, \label{X+prime}
\end{eqnarray}
with all other fields unchanged. This can also be expressed in terms of the superfields:  \begin{eqnarray}
\ba_i^{\prime} &=& \ba_{i} + H_{ijk}(\bx) ({\bm \varepsilon}_0 \bx^j) \bbd_1 \bx^k.
\label{tranformation2}
\end{eqnarray}
With this change of variables, the FHGD-BV symplectic form 
\eqref{FHGDBVsymplectic} becomes
\begin{eqnarray}
\omega_{BV}
&=& 
\int_{T[1]\Sigma} d^2 \sigma d^2 \theta \, \delta \bx^{i} \delta \ba_i^{\prime}.\label{newBV}
\end{eqnarray}
Below, there will be a more involved calculation which we will leave to the reader. This one we will prove in detail now:  We need to show that 
\beq \int_{T[1]\Sigma} d^2 \sigma d^2 \theta \, \delta \bx^{i} \delta \left[H_{ijk}(\bx) ({\bm \varepsilon}_0 \bx^j) \bbd_1 \bx^k \right] =  \int_{T[1]\Sigma} d^2 \sigma d^2 \theta \,
 \tfrac{1}{2} 
H_{ijk}(\bx) \bbd_1 \bx^i \delta \bx^j \delta \bx^k. \label{show}
\eeq 
Applying $\delta$ on the left on the term within the bracket leads to the following three terms:
\begin{eqnarray}
\int_{T[1]\Sigma} d^2 \sigma d^2 \theta \, \delta \bx^{i} \delta \left[H_{ijk}(\bx) ({\bm \varepsilon}_0 \bx^j) \bbd_1 \bx^k \right] &=&\int_{T[1]\Sigma} d^2 \sigma d^2 \theta \, 
 \delta \bx^{i} \left[H_{ijk,l}(\bx) \delta \bx^{l} ({\bm \varepsilon}_0 \bx^j) \bbd_1 \bx^k \right. \nonumber \\
 && \!\!\!\! \!\!\!\! \!\!\!\! \!\!\!\! \!\!\!\! \!\!\!\! \!\!\!\! \!\!\!\! \!\!\!\! \!\!\!\! \!\!\!\! \!\!\!\! \!\!\!\! \!\!\!\!\left. + H_{ijk}(\bx) ({\bm \varepsilon}_0 \delta \bx^j) \bbd_1 \bx^k - H_{ijk}(\bx) ({\bm \varepsilon}_0 \bx^j) \bbd_1 \delta \bx^k\right] ,
\end{eqnarray}
where we made use of the fact that $\delta$ commutes with ${\bm \varepsilon}_0$ and anticommutes with $\bbd_1$. Now a partial integration along the spatial $S^1$ (recall $\Sigma = S^1 \times S^1$) permits to rewrite the last term as follows:
\beq  \int_{T[1]\Sigma} d^2 \sigma d^2 \theta \, \delta \bx^{i}H_{ijk}(\bx) ({\bm \varepsilon}_0 \bx^j) \bbd_1 \delta \bx^k = \tfrac{1}{2}\int_{T[1]\Sigma} d^2 \sigma d^2 \theta \,  \bbd_1\left[ H_{ijk}(\bx) ({\bm \varepsilon}_0 \bx^j) \right] \delta \bx^{i}\delta \bx^k.
\eeq 
It now remains to use $\bbd_1\left[ H_{ijk}(\bx) {\bm \varepsilon}_0 \bx^j \right] = H_{ijk,l}(\bx) (\bbd_1 \bx^l){\bm \varepsilon}_0 \bx^j+  H_{ijk}(\bx) {\bm \varepsilon}_0 \bbd_1\bx^j$ on the right-hand side of this equation, to apply the identity 
\beq  \int_{T[1]\Sigma} d^2 \sigma d^2 \theta \: \alpha =   
\int_{T[1]\Sigma} d^2 \sigma d^2 \theta \: {\bm \varepsilon}_0 (\alpha ), \eeq
which is valid for every integrand $\alpha \in C^\infty({T[1]\Sigma})$, to the right-hand side of \eqref{show}, and to collect all the terms in the latter equation after use of the Leibniz rule for 
${\bm \varepsilon}_0$; all terms then cancel against one another on the nose, except for those containing 
derivatives of $H$, but which are seen to vanish precisely due to $\rd H=0$.

%
%

Implementing \eqref{tranformation2} to the BV action \eqref{superFHGDBVaction}, a similar calculation shows that 
it is transformed into 
\begin{eqnarray}
\Sbvbbd
&=& \int_{T[1]\Sigma}
d^2 \sigma d^2 \theta \ \left[
\ba_i^{\prime} \bbd \bx^i 
+  \tfrac{1}{2} \pi^{ij}(\bx) \ba_i^{\prime} \ba_j^{\prime}\right] + \frac{1}{3!} \int_{T[1]N} H_{ijk}(\bx) \bbd \bx^i \bbd \bx^j \bbd \bx^k 
\nonumber \\
&& \!\!\!\! \!\!\!\! \!\!\!\! \!\!\!\! \!\!\!\!+ \int_{T[1]\Sigma}
d^2 \sigma d^2 \theta \ \left[\tfrac{1}{2} \pi^{il} \pi^{jm} H_{klm}(\bx)
\ba_i^{\prime} \ba_j^{\prime} {\bm \varepsilon}_1 \bx^{k} 
- \pi^{ij} H_{jkl}(\bx) \ba_i^{\prime} ({\bm \varepsilon}_0 \bx^k) \bbd_1 \bx^l
\right] \label{superfieldBVaction3}
. 
\end{eqnarray}
In terms of the component fields, this is even simpler: Implementing \eqref{X+prime} into \eqref{FHGDcomponentBVaction}, there is only one term that changes. After slightly reorganizing terms, one obtains: 
\begin{eqnarray}
\Sbvbbd&=&
\int_{\Sigma} d^2 \sigma \left(
- A_{0i} \partial_1 \xzero^i + A_{1i} \partial_0 \xzero^i 
- \pi^{ij}(\xzero) A_{0i} A_{1j} \right)
+ \int_{N} H +
\nonumber \\ 
&&
 \int_{\Sigma} d^2 \sigma \left[
- \pi^{ij}(\xzero) \xzero^{+\prime}_{10 i} c_j
+ A_0^{+k} (\partial_1 c_i + f_k^{ij} A_{1i} c_j- \pi^{ij} H_{ikl} F_1^l c_j)
\right.
\nonumber \\ 
&&\qquad - A_1^{+k} (\partial_0 c_i + f_k^{ij} A_{0 i} c_j)
+ c_{10}^{+k} \tfrac{1}{2} f_k^{ij}
c_i c_j 
\left.
+ \tfrac{1}{2} f_k^{ij}{}_{,n} A_0^{+n} A_1^{+k} c_i c_j
\right],
\label{superfieldBVaction3components}
\end{eqnarray}
where 
$F_1^i \equiv \partial_1 X^i + \pi^{ij} A_{1j}$ is an on-shell vanishing contribution, see \eqref{eomdX}.

As anticipated, even after bringing $\omega_{BV}^{(FHGD)}$ into Darboux form \eqref{newBV}, there still is a problem with covariance of the BV functional. 
Correspondingly, the BV action \eqref{superfieldBVaction3} does not agree with the covariant result \eqref{superfieldBVaction} (or, equivalently, neither does \eqref{superfieldBVaction3components} with \eqref{supercoordBVaction}).

We thus now will apply the lesson we learned from the second toy model and the lemma we proved in the last section. To find the corresponding $\Lambda$ that one needs to choose here within Lemma \ref{lemmact}, we compare the BRST operator $s'$ that one obtains from \eqref{superfieldBVaction3components} with the $\mathrm{Diff}(\Sigma)$-covariant one $s$ that results from the gauge transformations (and that underlies the BV functional \eqref{supercoordBVaction}). There is a difference only in the action on the gauge fields $A_i$. On the one hand, we have (see \eqref{superfieldBVaction3components})
\begin{eqnarray}
s' A_{0i}&=& \partial_1 c_0 + f_k^{ij} A_{0i} c_j , \\
s' A_{1i}&=& \partial_1 c_i + f_k^{ij} A_{1i}c_j - \pi^{ij} H_{ikl} F_1^l  c_j ,
\end{eqnarray}
while on the other hand (see \eqref{gaugetransformation02a})
\begin{eqnarray}
s A_{0i}&=& \partial_1 c_0 + f_k^{ij} A_{0i}c_j - \tfrac{1}{2}\pi^{ij} H_{ikl} F_0^l  c_j , \\
s A_{1i}&=& \partial_1 c_i + f_k^{ij} A_{1i}c_j -\tfrac{1}{2} \pi^{ij} H_{ikl} F_1^l  c_j .
\end{eqnarray}
Here $F^i$ is the functional derivative of the classical action with respect to the field $A_i$, see \eqref{eomdX}. 
This comparison suggests the following choice: 
\begin{eqnarray}
\Lambda &=& \int_{\Sigma} d^2 \sigma \,
\tfrac{1}{2} \Theta^i_{jk}(X) A_0^{+j} A_1^{+k} c_i \, ,
\label{LambdainHPSM}
\end{eqnarray}
where $\Theta_{jk}^i \equiv \pi^{il} H_{jkl}$ is the torsion associated to the HPSM, see \eqref{torsionpiH}. Note that this has to be applied to the primed fields, i.e.\ the fields after the first transformation that permitted to put $\omega_{BV}^{(FHGD)}$ into Darboux form \eqref{newBV}. In the component field notation, this effected only $X_{10i}^+$, in the superfield notation $\ba_i$---for the notation, we chose to put the prime only on the fields that are changed, but this is to be kept in mind in what follows. For example, if one rewrites \eqref{LambdainHPSM} in terms of the superfields, it now should contain primes as follows: 
\begin{eqnarray}
\Lambda &=& 
- \int_{\Sigma} d^2 \sigma d^2 \theta \,
\tfrac{1}{2} \Theta^i_{jk}(\bx) ({\bm \varepsilon}_0{\bx^j})({\bm \varepsilon}_1{\bx^k})
\ba_i^{\prime}.
\label{LambdainHPSMsuper}
\end{eqnarray}

One now has to calculate the exponential $\exp({H_\Lambda})$ of the Hamiltonian vector field of $\Lambda$. It turns out that the exponential terminates already after the first term in this case. 
The BV canonical transformation generated in this way is simply of the form
\begin{eqnarray}
X^{\prime\prime i} &=& X^{i},
\label{change31}
\\
A^{+\prime\prime i} &=& A^{+i},
\label{change32}
\\
c_{10}^{+\prime\prime i} &=& c_{10}^{+i} + \frac{1}{2} \Theta_{jk}^i A_0^{+j} A_1^{+k},
\label{change33}
\\
c^{\prime\prime}_i &=& c_{i},
\label{change34}
\\
A_{0i}^{\prime\prime} &=& A_{0i} - \frac{1}{2} \Theta^j_{ki} A_0^{+k} c_j,
\label{change35}
\\
A_{1i}^{\prime\prime} & = & A_{1i} + \frac{1}{2} \Theta^j_{ki} A_1^{+k} c_j,
\label{change36}
\\
X_{10i}^{+\prime\prime} &=& X_{10i}^{+\prime} + \frac{1}{2} \Theta^j_{kl,i} A_0^{+k} A_1^{+l} c_j.
\label{change37}
\end{eqnarray}
If combined with the previous transformation, see \eqref{X+prime}, the last equation turns into
\beq X_{10i}^{+\prime\prime} = X_{10i}^{+} - H_{ijk} A_0^{+k} \partial_1 X^j + \frac{1}{2} \Theta^j_{kl,i} A_0^{+k} A_1^{+l} c_j \, .
\eeq 
The canonical transformations \eqref{change31}--\eqref{change37} can be also expressed in terms of superfields:
\begin{eqnarray}
\bx^{\prime\prime i} &=& \bx^i + \frac{1}{2} \Theta_{jk}^i(\bx) 
({\bm \varepsilon}_0 \bx^j) ({\bm \varepsilon}_1 \bx^k),
\label{changesuper1}
\\
\ba_{i}^{\prime\prime} 
&=& \ba_{i}^{\prime} 
+ \frac{1}{2} \Theta^j_{ki}(\bx) [- ({\bm \varepsilon}_0 \bx^k) (1- {\bm \varepsilon}_1) \ba_{j}^{\prime} + ({\bm \varepsilon}_1 \bx^k) (1 - {\bm \varepsilon}_0) \ba_{j}^{\prime}]
\nonumber \\ &&
- \frac{1}{2} (\partial_i \Theta^j_{kl} + \partial_l \Theta^j_{ki} - \partial_k \Theta^j_{li})(\bx) ({\bm \varepsilon}_0 \bx^k) ({\bm \varepsilon}_1 \bx^l) \ba_{j}^{\prime}. 
\label{changesuper2}
\end{eqnarray}

A direct calculation, which we now leave to the reader, establishes that expressing $S_{BV}^{(FHGD)}$ in terms of these fields, one indeed finds the covariant result $S_{BV}$ of \cite{Ikeda-Strobl}, 
\eqref{superfieldBVaction} or, equivalently, \eqref{supercoordBVaction}. This corresponds to the application of the exponential \eqref{expH}  to \eqref{superfieldBVaction3} or, equivalently, to \eqref{superfieldBVaction3components} .

\section{The FHGD formalism with $G$-covariance}
\label{secGeneral}
In this final section we want to generalize and formalize the procedure of restoring $G$-covariance in the Fisch--Henneaux--Grigoriev--Damgaard formulas---or some possibly other given, non-covariant BV extension of a gauge theory. Here $G$ is supposed to be the group of diffeomorphisms of $d$-dimensional spacetime, still denoted by $\Sigma$, or one of its subgroups, like, in particular, the group of Lorentz transformations for the case that $\Sigma$ is equipped with a Minkowski metric---but in principle one can also consider an arbitrary group $G$ acting on the space of fields.

Let us denote some initially given BV functional simply by $S$ here. We assume that the BV symplectic form is Darboux and decompose $S$ according to the polynomial degree ``$\mom$'' of  BV momenta (the antifields of the theory):\footnote{Before we used the notation $S^{(k)}$ for $S_k$. Note also that the polynomial degree in the antifields must not be confused with the antighost or antifield number that is used sometimes in literature and which is defined in a different way.} 
\beq S = \sum_{k=0}^\infty S_k \, . \label{decomp}\eeq 
So, in particular, $\mom(S_k) = k$ and $S_0$ is the classical action. $S$ is assumed to satisfy the classical master equation, 
\beq (S,S)=0 , \label{master1}
\eeq 
but it is not necessarily $G$-invariant. We may think of $S_{BV}^{FHGD}$ in the context of $S$, but in principle $S$ can be any initially given BV extension. 

Let us denote the functions (or local functionals) on the space of BV-fields of the theory by $\V= \oplus_{k=0}^\infty \V_k$. Evidently,  $S_k$ is precisely the component of $S$ inside $\V_k$. The BV bracket $( \cdot , \cdot )$ is homogeneous with respect to the degree $\mom$, decreasing it by one. Therefore, upon usage of \eqref{decomp}, we may decompose \eqref{master1} into an infinite set of coupled equations, the lowest one of which has again the simple form  $(S_0,S_0)=0$. By means of this equation together with the Leibniz rule of the BV bracket, the BV Hamiltonian vector field of the classical action $S_0$, 
\beq  \rd_0 := (S_0, \cdot ) \,  , \label{d0S0}
\eeq 
equips $\V$ with the structure of a complex:
\beq \V_0  \stackrel{\rd_0}{\longleftarrow} \V_1 \stackrel{\rd_0}{\longleftarrow} \V_2 \stackrel{\rd_0}{\longleftarrow} \V_3 \stackrel{\rd_0}{\longleftarrow} \ldots \label{complex}
\eeq

\subsection{Covariance in BV and the associated double complex}

We assume that there is a group $G$ acting on the (possibly non-linear) space of fields, which thus induces a (linear) $G$-action on the vector space $\V$. Most important in our context is the case where $G$ is the group of diffeomorphisms of spacetime, \eqref{Diff}, or an appropriate subgroup thereof which preserves background structures on $\Sigma$. Such a $G$-action arises if the fields of the theory,  including the ghosts and antifields, are naturally differential forms on $\Sigma$ (like it is the case for the HPSM), or also tensor fields (like a pseudo-Riemannian metric in gravity) or tensor densities. We will actually content ourselves here with an infinitesimal $G$-invariance, i.e.\ consider invariance with respect to 
\beq  \mathfrak{g} := \mathrm{Lie}(G).
\eeq 
In the cases of main interest then, elements of $\mathfrak{g}$ are vector fields $v$ on $\Sigma$ and their action on the fields is given by the corresponding Lie derivative. 

Given a $G$- or $\mathfrak{g}$-action on $\V$, we can consider the appropriate action Lie algebroid $E := \V \times \g$ and, after shifting the degrees of its fibers $\g$, introduce a BRST like operator or odd vector field $\rd_Q$ \`a la Vaintrob \cite{Vaintrob} on  $E[1] =  \V \times \g[1]$, which satisfies $\rd_Q^2 = \tfrac{1}{2}[\rd_Q,\rd_Q] = 0$ (the square bracket denotes the graded commutator). $\rd_Q$ carries a degree independent of the previous ones, which we denote by $\gdeg$: 
$\gdeg (\rd_Q) = 1$. 

Now it is decisive that, by construction, the classical action $S_0$ is $\mathfrak{g}$-invariant, 
\beq \rd_Q S_0 = 0. \label{dQS0}\eeq 
Let us assume furthermore that the BV symplectic form is $G$-invariant and that $\rd_Q$ does not change the polynomial degree $\mom$ of the antifields. This implies that $\rd_Q$ and $\rd_0$ are graded commutative, $[\rd_Q,\rd_0]=0$: Indeed, by the above assumption, $\rd_Q$ goes through the BV-bracket in the definition \eqref{d0S0} of $\rd_0$ and the commutativity then follows directly from \eqref{dQS0}. Thus we find that the \eqref{complex}  extends naturally into a bicomplex as follows:
\[
\xymatrix{
\vdots & \vdots & \vdots & \vdots & \\
\V_0 \otimes \wedge^2 \g^* \ar[u]_{\rd_Q} & \V_1 \otimes \wedge^2 \g^* \ar[l]_{\rd_0} \ar[u]_{\rd_Q} & \V_2 \otimes \wedge^2 \g^* \ar[l]_{\rd_0} \ar[u]_{\rd_Q} & \V_3 \otimes \wedge^2 \g^* \ar[l]_{\rd_0} \ar[u]_{\rd_Q} & \ar[l]_-{\rd_0} \ldots \\
\V_0 \otimes \g^* \ar[u]_{\rd_Q} & \V_1 \otimes \g^* \ar[l]_{\rd_0} \ar[u]_{\rd_Q} & \V_2 \otimes \g^* \ar[l]_{\rd_0} \ar[u]_{\rd_Q} & \V_3 \otimes \g^* \ar[l]_{\rd_0} \ar[u]_{\rd_Q} & \ar[l]_-{\rd_0} \ldots \\
\V_0 \ar[u]_{\rd_Q} & \V_1 \ar[u]_{\rd_Q} \ar[l]_{\rd_0}  & \V_2 \ar[u]_{\rd_Q} \ar[l]_{\rd_0} & \V_3 \ar[u]_{\rd_Q} \ar[l]_{\rd_0} & \ar[l]_-{\rd_0} \ldots
}
\]

It is advisable, moreover, to include the ghosts of the $\g$-action and their then-to-be-introduced antifields in the BV formulation of the gauge theory. Extending the BV symplectic form  appropriately, and denoting the corresponding brackets still in the same way, we thus assume that also the differential $\rd_Q$ has a Hamiltonian charge,
\beq \rd_Q = (Q, \cdot ) \, , \label{dQ}
\eeq
which now also permits to rewrite \eqref{dQS0} as 
\beq (Q,S_0)=0 . \label{QS0} 
\eeq  Such a situation arises, for example, when regarding the BV phase space as a shifted cotangent bundle over the action Lie algebroid $\V \times \g[1]$ (see \cite{Ikeda-Strobl,Kotov-Strobl1} for similar constructions).

\subsubsection{The two-dimensional example with diffeomorphisms}\label{sec:diff}
Let us illustrate the setting at our example with the PSM twisted by a 2-form $B$, where the group $G$ consists of all diffeomorphisms of $\Sigma$, which in turn is two-dimensional and which we assume to be compact without boundary in this subsection. The $\g$-ghosts,  $v \in C^\infty(\Sigma, \g[1])$, are thus essentially vector field valued fields 
\beq v^\mu = v^\mu(\sigma) \,  \label{v},
\eeq 
on $\Sigma$ of degree minus one. The space $\g^*$ dual to $\g$ can be identified with $\Omega^1(\Sigma) \otimes \Omega^2(\Sigma)$:  for an element $\alpha \in \g^*$ and $w \in \g$ the value $\alpha(w)$ is obtained by first contracting the vector field $w$ with the first entry of $\alpha$ and then integrating the resulting volume form over $\Sigma$. We thus consider the antifields $v^*$, the momenta of the  $\g$-ghosts, 
as $\Omega^1(\Sigma)$-valued volume forms on $\Sigma$ of degree plus two: 
\beq v^*_\mu = \tfrac{1}{2}v^*_{\mu\vert \nu \rho}(\sigma) \theta^\nu \theta^\rho \, .\label{v*}
\eeq 
The situation is a bit particular in a situation  where, as here, the Lie algebra $\g$ consists of sections over the same base space as the one used for the fields. To avoid an overly complicated notation, we thus chose a partial abstract index notation in \eqref{v} and \eqref{v*} (similar to the fields of the HPSM, in fact, where the abstract index referred a target space). In this notation, the extension of the BV symplectic form that now governs the $\g$BV double complex takes the following form
\begin{eqnarray}
\omega = \omega_{BV} 
- \int_{T[1]\Sigma} d^2 \sigma d^2 \theta \, \delta v^{\mu} \, \delta v^*_{\mu}. \label{extension}
\end{eqnarray}

The fields of the twisted PSM combine into differential forms on $\Sigma$ and there is a natural action of $\g$ on them by Lie derivatives using $\calL_{v} = [\uv, \rd]$ with $\uv \equiv v^\mu  \frac{\partial}{\partial \theta^{\mu}}$. On the other hand, the new ghosts \eqref{v} are vector fields and thus $(\calL_{v} v)^\mu = v^\nu \partial_\nu v^\mu$. The generator $Q$ of diffeomorphisms for the twisted PSM then takes the form
\begin{eqnarray}
Q &=&  \int_{T[1]\Sigma}
d^2 \sigma d^2 \theta \left[-\ba_i \calL_{v} \bx^i + (\calL_{v} v)^\mu \, v^*_\mu \right] \, ,
\label{BVQdiffeo}
\end{eqnarray}
where we used the superfield notation \eqref{2Dsuperfield1}, \eqref{2Dsuperfield2}. Decomposition  into homogeneous parts of the fields yields, more explicitly, 
\begin{eqnarray}
Q &=&\int_{T[1]\Sigma}
d^2 \sigma d^2 \theta \left[ - \ucplus^i \calL_{v} \uc_i + \uA_i \calL_{v} \uAplus^i - \uXplus_i \calL_{v} X^i + (\calL_{v} v)^\mu \, v^*_\mu  \right] \, .
\label{BVQdiffeo2}
\end{eqnarray}
Being the charge for the Vaintrob' operator of an action Lie algebroid, it is clear that it squares to zero: 
\begin{eqnarray}
(Q, Q) =0 \, ,
\end{eqnarray} 
which one also easily verifies directly. Note, on the other hand, that for every vector field $w$ and volume form $\alpha$ one has  $\int_\Sigma \calL_{w} \alpha = \int_\Sigma \iota_{w} (\rd \alpha) = 0$, since $\Sigma$ was assumed to not have a boundary and the contraction with a vector field of any differential form has vanishing top degree. Due to the Leibniz rule satisfied both by \eqref{dQ} and the Lie derivative, we thus conclude also that \eqref{QS0} holds true here, where $S_0$ is the classical action \eqref{BPSM}---again this can be as well established by a direct calculation certainly.

\subsubsection{Example of Lorentz transformations in $d$ dimensions}
It is illustrative to also consider the case of Lorentz transformations. Let $\Sigma = \R^d$ equipped with a Lorentzian metric $\eta = \tfrac{1}{2} \eta_{\mu \nu} d \sigma^\mu d \sigma^\nu$. All fields are now assumed to satisfy appropriate fall-off conditions such that integrals are well-defined. The action of the Lorentz group on fields living on the Minkowski space is defined as usually; for example, on a scalar field $\varphi$, a 1-form field $V$, and a covariant 2-tensor field $T$, the $\g$-transformations take the form
\begin{eqnarray}
\delta_w \varphi \equiv  -i w^{\lambda\rho} M_{\lambda\rho} \cdot \varphi &=& w^{\lambda\rho} \sigma_{\lambda} \partial_{\rho} \varphi,
\\
\delta_w V_{\mu} \equiv -i w^{\lambda\rho} M_{\lambda\rho}\cdot  V_{\mu} &=& w^{\lambda\rho} \sigma_{\lambda} \partial_{\rho} V_{\mu} + w_{\mu}{}^{\rho} V_{\rho},
\\
\delta_w T_{\mu\nu} \equiv -i w^{\lambda\rho} M_{\lambda\rho} \cdot T_{\mu\nu} &=& w^{\lambda\rho} \sigma_{\lambda} \partial_{\rho} T_{\mu\nu} + w_{\mu}{}^{\rho} T_{\rho\nu} + w_{\nu}{}^{\rho} T_{\mu\rho},
\end{eqnarray}
where $M_{\lambda\rho}$ denotes the generators of Lorentz transformations and $w_{\mu\nu} = - w_{\nu\mu}$ is a constant antisymmetric tensor parametrizing the transformation. Declaring $w_{\mu\nu}$ to become global ghosts on $\Sigma$, now the extension of the BV symplectic form becomes 
\beq \omega := \omega_{BV} + \delta w^{\mu\nu} \delta w^*_{\mu\nu} \, , \eeq 
where, in the second term, there is no integration over $\Sigma$ certainly. In addition to the original BV brackets of the fields, the only non-vanishing new brackets are \begin{eqnarray}
(w^{\mu\nu}, w^*_{\lambda\rho}) = \frac{1}{2} (\delta^{\mu}{}_{\lambda} \delta^{\nu}{}_{\rho} - \delta^{\mu}{}_{\rho} \delta^{\nu}{}_{\lambda}).
\end{eqnarray}
For concreteness, let us assume that the BV fields before the extension form a tower of differential forms $\ba_i$ and $\bx^i$ of all possible form degrees---with $\Sigma$ being two-dimensional, $d=2$, or not---such that 
\begin{eqnarray}
\omega_{BV}
&=& 
\int_{T[1]\Sigma} d^d \sigma d^d \theta \,( \delta \bx^{i} \delta \ba_i).\label{newBVd}
\end{eqnarray} Then the charge $Q$ for the Lorentz transformation looks as follows:
\begin{eqnarray}
Q &=& i \int_{T[1]\Sigma}
d^d \sigma d^d \theta \left(\ba_i w^{\lambda\rho} M_{\lambda\rho} \cdot \bx^i \right)
- w^*_{\mu}{}^{\lambda} w^{\mu}{}_{\nu} w^{\nu}{}_{\lambda} \, . \label{LorentzQ}
\end{eqnarray}

The group of Lorentz transformations is a subgroup of the group of diffeomorphisms of $\Sigma$. So, there should be a relation between the generators \eqref{BVQdiffeo} and \eqref{LorentzQ}---
in arbitrary dimensions of $d$, if we reinterpret the formulas of Sec.\ \ref{sec:diff} correspondingly, with $v_\mu^*$ being $d$-forms (due to the non-compactness of $\Sigma$, these fields are now also required to be of compact support or with appropriate fall-off conditions at infinity so that the integral \eqref{BVQdiffeo}  converges). 

Indeed, it is not difficult to verify that the formulas \eqref{newBVd} and \eqref{LorentzQ} follow from 
\eqref{extension} and \eqref{BVQdiffeo}, respectively, upon a (generalized or odd) coisotropic reduction with respect to 
\beq v^{\mu}(\sigma) := - w^{\mu\nu} \sigma_{\nu} \, .  \eeq
Hereby, it is possible to identify the antifield $w^*_{\mu\nu}$ of $w^{\mu\nu}$ with 
\begin{eqnarray}
w^*_{\mu\nu}
&=&
- \tfrac{1}{2} \int_{\Sigma}
d^d \sigma d^d\theta \ (v^*_{\mu} \sigma_{\nu} - v^*_{\nu} \sigma_{\mu} ).
\label{antivantiw}
\end{eqnarray}
Within this generalized reduction process, \eqref{antivantiw} serves as  a ``gauge condition'', which makes the equivalence of \eqref{LorentzQ} with the reduction of  \eqref{BVQdiffeo} transparent.

\subsection{Recursive procedure}
We now address the construction of a modified BV extension $S'$, which again satisfies the master equation $(S',S')=0$, but now is in addition $\g$-invariant: \beq \rd_Q S' = 0 . \label{dQS'}\eeq
Under fairly general conditions \cite{Henneaux-Fisch}, we may assume that $S$ is related to $S'$ by means of a canonical transformation; implementing the latter by means of the exponential of a Hamiltonian vector field, as in \eqref{expH}, we have\footnote{Instead of changing the BV action functional from $S$ to $S'$, we can in principle also modify the $G$-action: $(Q',S)=0$ is equivalent to $(Q,S')=0$ if one twists the charge $Q$ inversely to \eqref{S'S}, i.e.\ $Q' := \exp(-H_{\Lambda'}) \cdot Q$. The formulas determining the search for $\Lambda'$---or $\Lambda$, introduced below---are not effected in this way. }
\beq S' = \exp(H_{\Lambda'}) \cdot S,  \label{S'S}
\eeq 
with $H_{\Lambda'} = (\Lambda',  \cdot )$. Due to the exponential function, the search for an adequate $\Lambda'$ is a highly non-linear problem. We remark in parenthesis that this transformation must not change the total degree of $S$; this implies that $H_{\Lambda'}$ is even and $\Lambda'$ is odd. 

In a first step, we also develop the searched-for function $\Lambda'$ according to the degree $\mom$:
\beq  \Lambda' = \sum_{l=0}^\infty \Lambda'_l \, . \label{Lambda'}
\eeq 
Since the bracket is homogeneous in this degree (decreasing it by one), the combination of the three equations \eqref{dQS'}, \eqref{S'S}, and \eqref{Lambda'} can be decomposed accordingly.  We display the lowest equations that one obtains in this way: For the degrees zero and one, one finds 
\bea \rd_Q \left[ \exp(H_{\Lambda'_1})S_0 + (\Lambda'_0,S_1) + \tfrac{1}{2} (\Lambda'_0,(\Lambda'_0,S_2)) + \ldots \right] &=& 0, \label{0}\\ \rd_Q \left[ ( {\Lambda'_2},S_0) +  \exp(H_{\Lambda'_1}) \cdot S_1 + (\Lambda'_0,S_2) + \tfrac{1}{2} (\Lambda'_0,(\Lambda'_0,S_3)) + \ldots \right] &=& 0 , \label{1}\eea
where the dots also contain further terms to all orders of $H_{\Lambda'_1} \equiv ({\Lambda'_1}, \cdot)$ as it is of degree zero, while for degree two one has
\bea
\!\!\! \rd_Q \left[ ( {\Lambda'_3},S_0) +  \tfrac{1}{2} (\Lambda'_2, (\Lambda'_2,S_0))+ ( {\Lambda'_2},S_1) + \exp(H_{\Lambda'_1}) S_2 + (\Lambda'_0,S_3) + 
\ldots \right] &=& 0 \, . \label{2}
\eea
If we set 
\beq \Lambda'_0 := 0 \, , \: \: \Lambda'_1 := 0 \, ,  \label{lambdaprimes}
\eeq 
then the first equation above, Eq.\ \eqref{0}, simply reduces to \eqref{dQS0}. Also the other equations contain a finite number of terms now only: At degrees one and two, for example, we obtain 
\bea 
&& \rd_Q \left[ S_1 + ( {\Lambda'_2},S_0) \right] = 0, \label{21} \\ 
&& \rd_Q \left[ S_2  +  ( {\Lambda'_2},S_1) + ( {\Lambda'_3},S_0)+\tfrac{1}{2} (\Lambda'_2, (\Lambda'_2,S_0)) \right] = 0 \, . \label{22} 
\eea
In general, at order $n$ one finds the still highly non-linear coupled system for the $\Lambda'$s
\begin{align}
\rd_Q \left[ S_n 
+ \sum_{m=1}^n\sum_{j=0}^m \sum_{(p_1(m),\ldots, p_j(m)) \in {\cal P}(m)} a_{p_1, p_2, \ldots, p_j} ({\Lambda'_{p_1(m)+1}},( {\Lambda'_{p_2(m)+1}}, \ldots, (\Lambda'_{p_j(m)+1},S_{n-m})\ldots )) \right] = 0 \, , \label{2n}
\end{align}
where $(p_1(m), p_2(m), \ldots, p_j(m)) \in {\cal P}(m)$ is an element of the set of partitions of the  natural number $m$ such that $p_1(m) \geq p_2(m) \geq \ldots \geq p_j(m)$ and the coefficient $a_{p_1, p_2, \ldots, p_j}$ is the one in front of $x_{p_1(m)}x_{p_2(m)}\cdots x_{p_j(m)} y_{n-m}$ that appears in the expansion of the function $(\sum_{k=0}^{\infty} y_k)\cdot \exp(\sum_{l=0}^{\infty} x_l)$.

Luckily, there is another way of tackling the problem, reducing it to a linear one of cohomological nature within a recursion. For this purpose, let us define a sequence of extensions $(S')_m$ starting with $(S')_0 = S$ and arriving at $S'$ in the limit,
\beq S' = \lim_{m \to \infty} (S')_m \, . \label{limit}\eeq
For this purpose we require for every $m \geq 1$
\beq (S')_m := \exp(H_{\Lambda_{m+1}}) \cdot \exp(H_{\Lambda_{m}}) \cdot \ldots \cdot \exp(H_{\Lambda_2}) \cdot  S  \, ,  \label{something}
\eeq 
where $\mom ({\Lambda_k}) = k$.  

For every $m \geq k$,  $S'_k = (S')_{m,k}$, where  $(S')_{m,k}$ denotes the degree $k$ component of $(S')_m$.  So for every fixed degree $l$, the redefinition of  $(S')_{m,l}$, that happens  for small values of $m$, stabilizes at some point and the limit \eqref{limit} is well-defined. The relation between $\Lambda$ and $\Lambda'$ in \eqref{S'S} is rather intricate, but for degree reasons $\Lambda$ is determined uniquely in terms of $\Lambda'$ and $S$, if \eqref{limit} is to hold true.  

Now we proceed by induction to determine $\Lambda$. Assume that $\rd_Q (S')_{m,k} = 0$ holds true for every $k \leq m$. Evidently this equation is satisfied for $m=0$ due to \eqref{dQS0}. Now we want to determine the conditions on $\Lambda_{m+2}$ such that this equation is satisfied also when $m$ is replaced by $m+1$. First we observe that  $\rd_Q (S')_{m+1,k} = 0$ holds true for all $k \leq m$ since $\exp(H_{\Lambda_{m+2}})$ acts as the identity on $\V_0 \oplus \V_1 \oplus \ldots \oplus \V_m$. Let us write out the condition $\rd_Q (S')_{m+1,m+1} = 0$; it is very simple,
\beq \rd_Q \left[(S')_{m,m+1} + (\Lambda_{m+2},S_0) \right] = 0 \, ,\label{condition1}
\eeq 
and can be rewritten  as the following condition on $\Lambda_{m+2}$:
\beq \rd_Q \rd_0 \Lambda_{m+2} = -\rd_Q (S')_{m,m+1} \, .
\eeq 
Denote by $C_Q$ the $\rd_Q$-cycles at  $\g$deg zero,
\beq (C_Q)_k :=  \{v \in \V_k \vert \rd_Q (v) = 0 \} \, , \eeq 
and define 
\beq \W_k := \V_k /  (C_Q)_k \ni [v] \, . \eeq
Then, due to the nature of a double complex, $\rd_0$ descends to a differential on $\W$ and the condition \eqref{condition1} can be rewritten as 
\beq \rd_0 [\Lambda_{m+2}] = -[ (S')_{m,m+1}] \, . \label{condition}\eeq  
Here, for any $\alpha \in \V_\bullet$, $[\alpha]$ denotes the corresponding element in the quotient complex $\W_\bullet$.

Let us recall now that $ (S')_{m,m+1}\equiv  \left[\exp(H_{\Lambda_{m+1}}) \cdot \exp(H_{\Lambda_{m}}) \cdot \ldots \cdot \exp(H_{\Lambda_2}) \cdot  S \right]_{m+1}$ and that by assumption all the $\Lambda_l$s entering this equation have been determined already in an earlier step. We also know that one has $(S')_{m,0}=S_0$ and 
\beq ((S')_{m},(S')_{m})=0 \, ;
\eeq 
Spelling out this last equation at $\mom$ equal to $m+1$, one obtains
\beq (S_0, (S')_{m,m+1})=  - \tfrac{1}{2} \sum_{l=1}^m ((S')_{m,l},(S')_{m,m+1-l}) \, . \label{rhs}
\eeq
Applying $\rd_Q$ to this equation,  we see that its right-hand side vanishes due to our recursion assumption, and we thus obtain
\beq \rd_Q \rd_0 (S')_{m,m+1} = 0 \qquad \Rightarrow \qquad \rd_0 [(S')_{m,m+1}] = 0 \, . 
\eeq 
This is the consistency condition following from \eqref{condition}; it is always satisfied here. 

We thus reduced the problem to a cohomological one in the complex 
\beq  \W_0  \stackrel{\rd_0}{\longleftarrow} \W_1 \stackrel{\rd_0}{\longleftarrow} \W_2 \stackrel{\rd_0}{\longleftarrow} \W_3 \stackrel{\rd_0}{\longleftarrow} \ldots. \label{complexW}
\eeq 
Down to earth, the transition from the complex $\V$ to the complex $\W$ means that within the redefinition procedure we may, both for $S'_l$ and $\Lambda_l$,  always drop $G$-covariant terms, putting them effectively to zero. This transition provides an important simplification: In general, $\rd_0 (S')_{m,m+1}$ will not vanish---it equals the r.h.s.\ of \eqref{rhs}---and thus certainly $(S')_{m,m+1}$ also will not be $\rd_0$-exact. However, for the first non-trivial choice of $m$, $m=0$, we showed in Section \ref{secChange} that the corresponding equivalence class $[(S')_{m,m+1}]$ \emph{inside} $\W_1$  is always $\rd_0$-exact and that thus the existence of $\Lambda_2$ satisfying \eqref{condition} with $m=0$ is guaranteed.

Let us remark that under the condition \eqref{lambdaprimes}, which is equivalent to $\Lambda_i = 0$ for $i=0,1$, the necessary and sufficient conditions for the existence of a $G$-covariant deformation are provided by the existence of $\Lambda'$ such that the coupled system \eqref{2n} holds true. 
In contrast, the above recursive procedure could lead at a particular step to a non-trivial cohomological element, which still could be overcome by changing the recursion at a previous step: indeed, modifying $\Lambda$ at a lower level by an element of non-trivial $\rd_0$-cohomology, a modification which is permitted at each step when solving \eqref{condition}, might change the cohomology class obtained at a later level. And, if this does not work and the above recursion procedure is obstructed also by such modifications, there still may be options to make it work by means of a non-trivial choice for $\Lambda'_0$ and $\Lambda'_1$: For example, as we learn from \eqref{0} and \eqref{1}, we might still solve the first of these two equations, but then change the second one decisively, where now even higher orders of $S$ enter the classes one obtains at lower orders.

Summing up, we reduced the search for a $G$-covariant modification of a given BV extension $S$ to a recursion of cohomological nature. We are not aware of results about the $\rd_0$-cohomology at different degrees in \eqref{complexW} that would guarantee existence of the $G$-invariant extension $S'$ in general. In practice, however, the improved FHGD procedure, where $S=S_{FHGD}$ and $\Lambda$ is determined recursively using \eqref{condition} so as to yield $S'$, defined as in 
 \eqref{something} and \eqref{limit}, should very often lead to covariant results. For the PSM (the HPSM with vanishing $H$ and $B$), the FHGD procedure yields a covariant BV functional on the nose. For the HPSM with a generic choice of $H$ and $B$, this is not the case, but the above procedure leads to the desired result already after the first step of the correction ($\Lambda_k:=0$ for all $k>2$).

\section*{Acknowledgments}
We are grateful to Maxim Grigoriev for discussions. N.I.~wants to express his gratitude to Universit\'e Lyon 1 and the Institut Camille Jordan for their hospitality during his stay in March 2020 and to the ESI for the possibility to present related work within the program ``Higher structures and field theories''.

This work was supported by the LABEX MILYON 
(ANR-10-LABX-0070) of Universit\'e de Lyon, within the program ``Investissements d'Avenir'' (ANR-11-IDEX-0007) operated by the French National Research Agency (ANR).


\end{document}